\documentclass[amsmath,amssymb,reprint,aip,jcp,longbibliography]{revtex4-1}

\usepackage{graphicx}% Include figure files
\usepackage{dcolumn}% Align table columns on decimal point
\usepackage{bm}% bold math
\usepackage{subfigure}
\usepackage{color}
\usepackage{natbib}
\usepackage{mathrsfs}
\usepackage{ccfonts}
\usepackage{xifthen}% provides \isempty test

\DeclareMathAlphabet{\mathpzc}{OT1}{pzc}{m}{it}

\newcommand{\bra}[1]{\ensuremath{\left\langle #1\right|}}
\newcommand{\ket}[1]{\ensuremath{\left|#1\right\rangle}}
\newcommand{\bigbraket}[2]{\left\langle {#1} \mathrel{\left | {\vphantom {#1 #2}} \right. \kern-\nulldelimiterspace} {#2} \right\rangle}

\renewcommand{\S}{\ensuremath{\mathbb{S}}}

\newcommand{\al}[3]{%
  \ifthenelse{\isempty{#3}}%
  {\alpha^{#2}_{#1}}%
  {\alpha^{#2}_{#1;#3}}%
}

\newcommand{\baralpha}{\bar{\alpha}}
\newcommand{\bal}[3]{%
  \ifthenelse{\isempty{#3}}%
  {\baralpha^{#2}_{#1}}%
  {\baralpha^{#2}_{#1;#3}}%
}

\newcommand{\tal}[3]{%
  \ifthenelse{\isempty{#1}}%
  {
    \ifthenelse{\isempty{#3}}%
               {\tilde{\alpha}^{#2}}%
               {\tilde{\alpha}^{#2}_{#3}}%
  }
  {
    \ifthenelse{\isempty{#3}}%
               {\tilde{\alpha}^{#2}_{#1}}%
               {\tilde{\alpha}^{#2}_{#1;#3}}%
  }
}

\begin{document}

\title{2D Stimulated Resonance Raman Spectroscopy of Molecules with
  Broadband X-ray Pulses}

\author{Jason D. Biggs}
\author{Yu Zhang}
\author{Daniel Healion}
\author{Shaul Mukamel}
\email{smukamel@uci.edu}

\affiliation{Department of Chemistry, University of California, Irvine\\Irvine, CA  92697-2025, USA.}%Lines break automatically or can be forced with
\date{\today}

% FWHM ~ 14.2eV (181 as)
%

\begin{abstract}
  Expressions for the two-dimensional Stimulated x-ray Raman
  Spectroscopy (2D-SXRS) signal obtained using attosecond x-ray pulses
  are derived. The 1D- and 2D-SXRS signals are calculated for
  trans-N-methyl acetamide (NMA) with broad bandwidth (FWHM
  $\simeq$14.2eV, 181 as) pulses tuned to the oxygen and nitrogen
  K-edges.  Crosspeaks in 2D signals reveal electronic Franck-Condon
  overlaps between valence orbitals and relaxed orbitals in
  the presence of the core hole.
 \end{abstract}
\maketitle

\section{Introduction}
\label{sec:introduction}

Driven by the development of highly intense attosecond sources of x-ray
radiation,\cite{marangos_introduction_2011} there is much interest in
mapping nonlinear spectroscopic techniques developed for the
NMR, IR and visible regimes to x-ray frequencies.\cite{popmintchev_attosecond_2010,adams_nonlinear_2003} These
techniques use sequences of laser pulses with well-defined inter-pulse
delays to probe the correlation between different nuclear spin,
vibrational, or electronic molecular eigenstates.  One-dimensional
stimulated Raman s-ray spectroscopy (1D-SXRS) has been proposed to
prepare and probe valence electronic wavepackets during a delay
between two pulses.\cite{tanaka_probing_2003} We present the two
dimensional stimulated x-ray Raman spectroscopy technique (2D-SXRS),
an extension of 1D-SXRS which may be used to probe molecular valence
electronic excitations in greater detail.  First we briefly survey the
optical domain, time-resolved, stimulated Raman experiments which inspire
these new x-ray techniques. We then will discuss the dynamics
probed by the time and frequency domain x-ray spectroscopies,
including 2D-SXRS.

Resonances are observed in optical or ultraviolet Raman spectroscopy
when the difference frequency $\omega_1 - \omega_2$ of the incident
($\omega_1$) and scattered ($\omega_2$) field modes match a
vibrational transition
frequency.\cite{mukamel_principles_1999,icors2010} This inelastic
scattering process can be either spontaneous or stimulated by a second
interaction with the applied field.
\cite{eckhardt_stimulated_1962,bloembergen_stimulated_1967,penzkofer_high_1979}
Femtosecond lasers with bandwidths comparable to molecular vibrational
frequencies (on the order of hundreds of wavenumbers), have made it
possible to excite vibrations
impulsively.\cite{mukamel_manipulation_1991,becker_femtosecond_1989,rosker_femtosecond_1986,weiner_three-pulse_1985}
The dynamics of a slow vibrational system are probed through a
perturbation that depends on its coupling to a fast electronic
system.\cite{dhar_time_1994,mukamel_manipulation_1991,biggs_using_2009}
Time-domain vibrational Raman spectroscopy with picosecond delays
between pulses was developed in the
seventies\cite{penzkofer_high_1979} and used to probe vibrational
dephasing.  The impulsive pump-probe
technique\cite{becker_femtosecond_1989,rosker_femtosecond_1986,weiner_three-pulse_1985,zewail_femtochemistry:_2000}
is one dimensional (1D) since a single delay time is controlled.
Selecting a specific intermediate electronic state by tuning the
central frequencies of the exciting pulses determines a unique
effective nuclear perturbation during the Raman process.
Multidimensional time-domain stimulated Raman spectroscopies extend
the stimulated Raman technique to multiple time evolution
periods. Fifth-order off-resonant stimulated Raman spectroscopy was
first proposed to examine the vibrational structure in
liquids.\cite{tanimura_two-dimensional_1993}

These nonlinear techniques were developed in parallel with
technological progress in the coherence and intensity of ultrafast lasers.  A new generation of very intense ultrafast x-ray sources
allow weak nonlinear x-ray transitions to be observed for the first
time despite the low cross-section and short lifetimes of
core-excitations.  X-ray free electron lasers
(XFELs)\cite{mcneil_x-ray_2010} convert the kinetic energy of a beam
of accelerated electrons to an electromagnetic field by passing the
beam through a magnetic grating.\cite{luchini_undulators_1990} The
Linac Coherent Light Source (LCLS) at
Stanford\cite{_lcls:_2000_book,emmap._first_2010} is a next generation
XFEL capable of generating extremely intense ($\sim 10^{13}\,$
photons) pulses with x-ray central frequencies
($560-2660$ eV),\cite{marangos_introduction_2011} and has already been
used to probe exotic forms of plasma\cite{vinko_creation_2012} and to
find the structure of biological samples from a succession of single
protein x-ray diffraction
snapshots.\cite{neutze_potential_2000,quiney_biomolecular_2011} The
source is intense enough to create hollow atoms by depopulating both
core orbitals for a resonant transition within the Auger decay time of
the core hole in atomic\cite{young_femtosecond_2010} and
molecular\cite{hoener_ultraintense_2010} systems.
One drawback of XFEL radiation is that the
coherence time is short ($<5$ fs),\cite{gutt_coherence_2012} a consequence of quantum
statistics in the electron bunch used to generate the pulse.
%One drawback of XFEL
%radiation is that the longitudinal autocorrelation function of the
%intense electric field decays very quickly, a consequence of quantum
%statistics in the electron bunch used to generate the
%pulse.\cite{gutt_coherence_2012}
Despite this drawback, the high intensity of the LCLS XFEL source make it the most likely candidate
for exciting nonlinear x-ray transitions. The technology needed to
create pulses with desired characteristics is rapidly improving; an
XFEL pumped x-ray laser was recently demonstrated, capable of
generating intense, temporally coherent x-ray pulses.\cite{rohringer_atomic_2012} Similar advances promise to
significantly extend the capability of the LCLS light source.

Pump-probe and stimulated Raman techniques were among the first
nonlinear experiments performed in the visible
regime.\cite{p.m._direct_1968,eckhardt_stimulated_1962,bloembergen_stimulated_1967,penzkofer_high_1979}
This paper focuses on x-ray analogues of multidimensional stimulated
Raman techniques, using soft or hard x-ray pulses with bandwidths
greater than 10 eV to create wave packets of electronic
excitations. The x-ray interaction involves single particle, field
driven, transitions between core and valence electronic orbitals and
the many-body valence response to a transiently created
core-hole.\cite{gadzuk_core_1987} Extensive
theoretical\cite{rohringer_x-ray_2007} and
experimental\cite{glownia_time-resolved_2010} work was applied toward using
this unique source as a probe in time-dependent experiments.  The low
cross-section of the x-ray matter interaction make IR and optical
pump/x-ray probe experiments more accessible than the all-x-ray
pump-probe technique.  IR pump/x-ray probe experiments were proposed
and simulated for diatomic
molecules.\cite{felicissimo_principles_2005,guimaraes_quantum_2005,guimaraes_pump-probe_2006,felicissimo_enhancement_2005}
Time-resolved optical pump/x-ray probe
experiments\cite{guimaraes_two-color_2004} have been performed at the
LCLS which highlight the experimental difficulties in synchronizing
pump and probe pulses.\cite{glownia_time-resolved_2010} All-x-ray
stimulated Raman experiments in which the pump and probe pulses both have x-ray
frequencies are a natural extension of these
techniques.  Frequency-domain resonant inelastic x-ray scattering
(RIXS) is a well established technique that probes the single-particle occupied and unoccupied density of
states around the resonant core by measuring the energy-resolved x-ray
radiation spontaneously emitted by a core-excited state
.\cite{gelmukhanov_resonant_1999,nordgren_soft_1989} An x-ray photon
is absorbed, exciting a core electron to an empty (virtual) orbital,
and an electron drops into the core-hole, emitting a photon.  Peaks in
the Fourier transform of the time-domain
1D-SXRS\cite{tanaka_coherent_2002} signal represent excitations of
valence electronic states generated through a core-excited
intermediate.  Ultrashort x-ray pulses are difficult to manipulate\cite{nugent_coherent_2010}
and the cross-section for core-excitation is lower than for
excitations at lower frequencies.  The feasibility of stimulated and coherent Raman experiments
using the LCLS was explored,\cite{patterson_resource_2010} and schemes for
tailoring the electron bunches to generate pairs of attosecond pulses
suited for stimulated x-ray Raman experiments have been
suggested.\cite{zholents_obtaining_2010}

Simulations\cite{kuleff_tracing_2007} and
experiments\cite{krausz_2010} suggest that electron correlation can drive
charge migration in ionized electronic systems. Thanks to the elemental
specificity of resonant x-ray excitation, it can be used as a perturbation
that is local in time and space, making nonlinear x-ray spectroscopies
attractive for measuring ultrafast dynamics with high spatial resolution.  Nonlinear effects in the
propagation of XUV pulses,\cite{reiter_route_2010} and atomic
attosecond electronic dynamics,\cite{tzallas_extreme-ultraviolet_2011}
were investigated in noble gases, and show that nonlinear effects
contribute to measurements using XFEL sources. The 1D-SXRS signal can be expressed as the time-dependent overlap of an impulsively excited valence electron doorway wavepacket with a stationary window created by the probe.
SXRS was proposed to investigate the electronic
properties of molecules,\cite{tanaka_coherent_2002} and applied to
conjugated $\pi$-bonded organic molecules\cite{tanaka_probing_2003}
and excitonic systems\cite{tanaka_time-resolved_2003} using a
tight-binding model Hamiltonian.  Several extensions and refinements
of 1D-SXRS were proposed, including the use of attosecond pulses to
prepare entangled particle-hole
states,\cite{mukamel_manipulating_2010} frequency domain coherent
anti-Stokes Raman spectroscopy with wide- and narrow-band pulses
\cite{rahav_manipulating_2009}, and a many-body Green's function-based
method suitable for calculating the SXRS of larger
systems.\cite{harbola_coherent_2009} X-ray spectra of molecules with
cores separated from each other at some distance were
discussed.\cite{schweigert_probing_2007}

This paper extends the formalism previously applied to
1D-SXRS to two dimensions (2D-SXRS) by adding one more pulse.
Expressions for the 1D-SXRS and 2D-SXRS signals are presented in
sections \ref{sec:1dsxrs} and \ref{sec:2dsxrs},
respectively. Simulations of the UV, 1D and 2D-SXRS signals for the nitrogen and oxygen K edges in
trans-NMA are reported in Sec. \ref{sec:results}, and their
significance is discussed in Sec. \ref{sec:concluding}.

\section{Ultrafast One Dimensional Stimulated Raman: 1D-SXRS}
\label{sec:1dsxrs}

\begin{figure*}[htbp]
  \centering
  \includegraphics[width=6in]{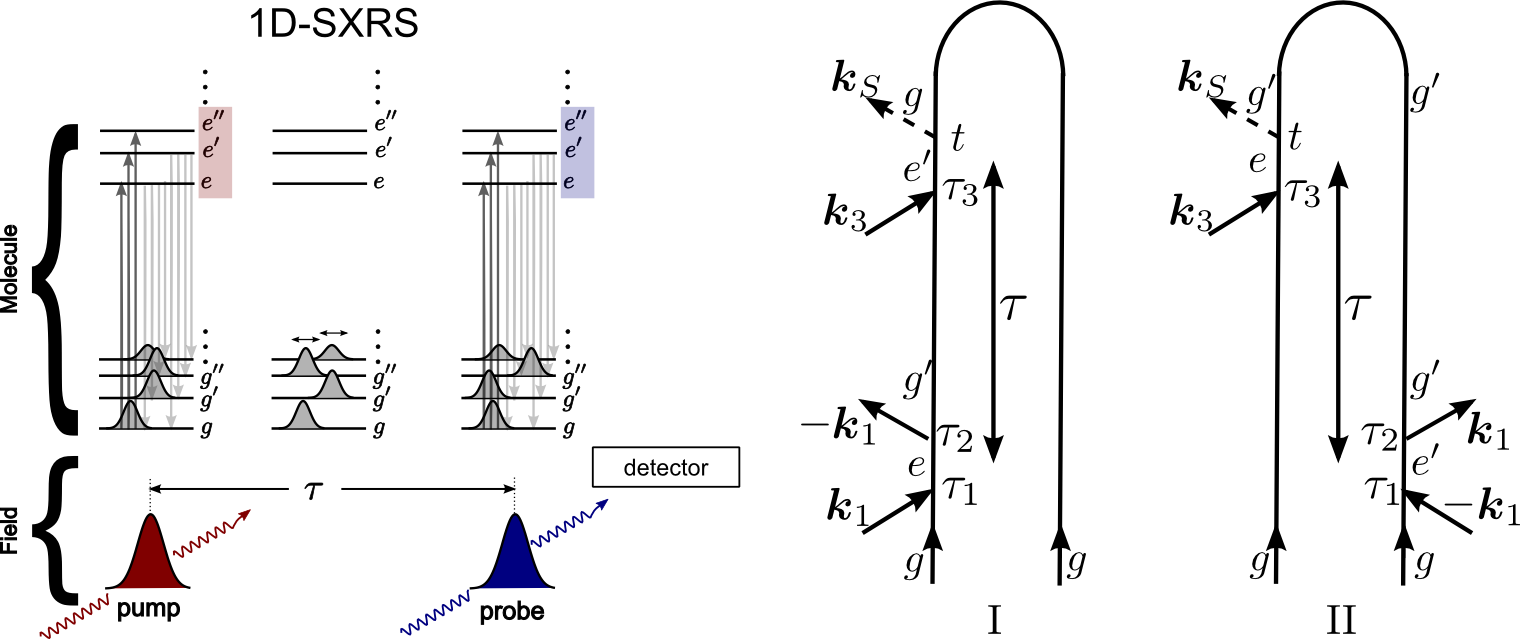}
  \label{fig:1Draman-diagram}
  \label{fig:energy-levels}
  \caption{\label{fig:1DRaman}Loop diagrams, pulse sequence,
    and core and valence energy levels for the 1D-SXRS signal.}
  \label{fig:1dsxrstechnique}
\end{figure*}

1D-SXRS is the simplest time-domain, x-ray Raman technique. The
electric field is represented as
\begin{equation}\begin{split}
  E(\mathbf{r},t) =& \mathbf{e}_1 \mathcal{E}_1(t)
  \exp(i \mathbf{k}_1 \cdot \mathbf{r} - i \omega_1 t) +
  \mathbf{e}_2 \mathcal{E}_2(t-\tau) \\ & \qquad \times
  \exp(i \mathbf{k}_2 \cdot \mathbf{r} - i \omega_2 (t-\tau)) + \textrm{c.c.}.
\end{split}\end{equation}
Here $\mathbf{k}_j$, $\omega_j$, $\mathbf{e}_j$ and $\mathcal{E}_j$
are the wavevector, carrier frequency, polarization vector, and complex
envelope of the $j$th pulse, respectively.  The system is excited by a pump pulse
and the transmission of the probe pulse is recorded after a delay
$\tau$ (see
Fig. \ref{fig:1dsxrstechnique}). \cite{tanaka_probing_2003,schweigert_probing_2007,healion_simulation_2011}
\par
Each pulse interacts with the system twice via a stimulated Raman
process. A core-electron is excited into an unoccupied orbital during
each pulse; the excited system then evolves for a short period before
a second interaction with the same pulse stimulates a valence electron
to destroy the core-hole, emitting an x-ray photon in the process.
The system is left in a coherent superposition of valence excited
states after interaction with the first x-ray pulse which interferes
with the amplitude of the Raman process with the second pulse, leading
to a change in its transmitted
intensity.\cite{patterson_resource_2010} The 1D-SXRS signal is defined
as the change in the transmission of the second pulse with or without
the pump pulse, as a function of the delay between the two pulses.  We
neglect nuclear motions,
further work will be needed to examine their effect on
SXRS.

We consider stimulated resonance Raman excitation at soft x-ray
wavelengths, which are much longer than the spatial extent of the
resonant core orbitals.  A phase dependent on the distance between
resonant cores will not enter into the response, as each field
interacts twice, canceling the phase factor of $\exp(\pm i
\mathbf{k}_j \cdot \mathbf{R}_n)$. Taking the delay
between pulses ($\tau$) to be long compared with the lifetime for
core-excited states ($<10$ fs for nitrogen and
oxygen),\cite{zschornack_handbook_2007} the signal will be dominated
by the ground-state bleach (GSB) contributions to the time-integrated
pump-probe signal (defined as the pump-induced change in the
absorption of the probe).

The closed-time path loop diagrams for the two interfering
contributions to the signal are shown in Fig
\ref{fig:1dsxrstechnique}.  These diagrams are similar to the
double-sided ladder diagrams traditionally used to describe nonlinear
optical spectroscopy,\cite{mukamel_principles_1999} with a few key
differences.  We read the loop diagram by starting with the ground
state on the lower left corner and moving upwards, describing forward
evolution (propagation of the ket).  Interactions with the field are
represented by arrows facing into (absorption of a photon) and out of
(emission of a photon) the diagram.  At the top of the loop we switch
to backward time evolution (propagation of the bra), and finally
arrive back at the ground state.  Thus we are able to work in Hilbert
space, rather than the higher dimensional Liouville space required
with the density matrix (for diagram rules see
Refs. \onlinecite{mukamel_chapter_2010} and
\onlinecite{biggs_coherent_2011}).  From Fig \ref{fig:1dsxrstechnique}
we get
\begin{widetext}
\begin{equation} \begin{split}\label{eq:1Da}
\S_{SXRS}(\tau) =& -\sum_{e,e',g'}V_{ge'} V_{e'g'} V_{g'e} V_{eg}\int_{-\infty }^{\infty }dt \int_{-\infty }^t d\tau_3 \int_{-\infty }^\infty d\tau_2 \int_{-\infty }^{\tau_2}d\tau_1  \mathcal{E}_2^*(t-\tau) \mathcal{E}_2(\tau_3-\tau) e^{-i \omega_2(\tau_3-t)}\\
& \times \bigg{[}\mathcal{E}_1^*(\tau_2) \mathcal{E}_1(\tau_1) \exp(-i \omega_{e'g}t+i \omega_{e'g'}\tau_3- i \omega_{eg'}\tau_2 +i \omega_{eg}\tau_1-i\omega_1(\tau_1-\tau_2))  \\
& \, \, \, \, \, \, \, \, + \mathcal{E}_1(\tau_2) \mathcal{E}_1^*(\tau_1) \exp(-i \omega_{eg'}t+i \omega_{eg}\tau_3+ i \omega_{e'g'}\tau_2 -i \omega_{e'g}\tau_1+i\omega_1(\tau_1-\tau_2)) \bigg{]} + \textrm{c.c.}
\end{split}\end{equation}\end{widetext}
Here \ket{g} is the electronic ground state, \ket{e} is a state with a
core hole, and \ket{g'} is a valence excited state (see
Fig. \ref{fig:1dsxrstechnique}).  Eq. \ref{eq:1Da} can be recast as
\begin{equation}\label{eq:sxrstime}
\S_{SXRS}(\tau) = \Re \left[\langle\alpha_2(\tau)\alpha_1(0) \rangle - \langle \alpha^\dagger_1(0)\alpha_2(\tau)\rangle\right]
\end{equation}
where
\begin{equation}\begin{split}
    \label{eq:13}
    \alpha_{j;g' g''} \equiv & -i \sum_e
    (\boldsymbol{e_j}\cdot\boldsymbol{V}_{g' e})
    (\boldsymbol{e_j}\cdot\boldsymbol{V}_{e g''})
    \int _{-\infty }^{\infty } d\tau _2 \int _{-\infty }^{\tau _2}
    d\tau _1 \\ & \times
    \mathcal{E}_j^*\left(\tau _2\right)\mathcal{E}_j\left(\tau _1\right)
     \times \exp \left(i \Delta^j_{eg'}\tau_2
    -i\Delta^j_{eg''}\tau_1
    \right)
\end{split}\end{equation}
is the effective polarizability weighted by the two-photon spectral
density of the $j^\textrm{th}$ ultrashort pulse.  Here $\Delta^j_{e
  \nu} \equiv \omega_j-\omega_{e \nu}+i \Gamma_{e}$ is the detuning
for the $\nu \rightarrow e$ transition, and $\Gamma_e$ is the inverse
of the core-hole lifetime.  In Eq. \ref{eq:13} we have included the
direction cosines between the pulse polarization vector and the
transition dipoles, as well as the core-excited lifetime
$\Gamma_e$. Eq. \ref{eq:13} can be recast in the frequency domain as
\begin{equation} \begin{split}
    \label{eq:13b}
    \alpha_{j;g' g''}=& \sum_e
    \frac{(\boldsymbol{e_j}\cdot\boldsymbol{V}_{g' e})(\boldsymbol{e_j}\cdot\boldsymbol{V}_{e g''})}{2\pi} \\ & \times
    \int _{-\infty }^{\infty } d\omega_2 \frac{\mathcal{E}_j^*\left(\omega_2\right)\mathcal{E}_j\left(\omega_2+\omega_{g'g''}\right)}{\omega_2+\Delta^j_{eg'}}
    \end{split}\end{equation}
\par
The first term in Eq. \ref{eq:sxrstime} (diagram I from Fig. \ref{fig:1DRaman}) can be viewed as a valence wavepacket $\alpha_1 \ket{\psi_0}$, created by
pulse 1, which propagates forward in time $\tau$ and overlaps with a window wavepacket $\bra{\psi_0}\alpha_2$ created by pulse 2.  The second term (diagram II) can be
viewed as a window wavepacket created by pulse 2 $\alpha_2 \ket{\psi_0}$ propagating backwards in time $-\tau$ to overlap with the doorway $\bra{\psi_0}\alpha_1$ created by pulse 1.

Hereafter we assume linearly polarized Gaussian pulses, with the  spectral envelope function
\begin{equation}\label{eq:ejgauss}
 \mathcal{E}_j(\omega) = \sqrt{2\pi} \sigma_j e^{-\sigma_j^2\omega^2/2-i \phi_j} ,
\end{equation}
where $\sigma_j$ is the temporal pulse width (equal to intensity FWHM
divided by $2 \sqrt{\ln 2}$).  The absolute phase $\phi_j$ of the
$j^\mathrm{th}$ pulse does not affect the signals considered here.
Inserting Eq. \ref{eq:ejgauss} in Eq. \ref{eq:13b}, the effective
polarizability becomes
\begin{equation}\begin{split}\label{eq:alpha}
  \alpha_{j;g' g''}=&-\frac{1}{2} \sum_e (\boldsymbol{e_j}\cdot\boldsymbol{V}_{g' e})(\boldsymbol{e_j}\cdot\boldsymbol{V}_{e g''})\mathcal{E}_j^*(\Delta^j_{eg'})\mathcal{E}_j(\Delta^j_{eg''})
    \\ & \times \left[i+\textrm{erfi}(-\sigma_j (\Delta^j_{eg'}+\Delta^j_{eg''})/2)\right].
\end{split}\end{equation}
where
\begin{equation}
\textrm{erfi}(x) = -\frac{2i}{\sqrt{\pi}} \int_0^{i x}dt e^{-t^2}
\end{equation}
is the imaginary error function (see Appendix \ref{sec:derivation} for
details).  The treatment so far (Eqs. \ref{eq:13}, \ref{eq:13b}, and
\ref{eq:alpha}) has assumed the molecule is oriented in the lab frame.
Of course, under typical experimental conditions in the gas or
condensed phases, the material will be an ensemble of randomly
oriented molecules.  All signals presented below are orientationally
averaged, assuming all incoming fields are polarized parallel to each
other, using the framework of
Ref. \onlinecite{andrews_three-dimensional_1977}.  The necessary tensor expressions
for the 2D-SXRS signal are given in Appendix \ref{sec:rotationalavg}.

The form for the effective polarizability given in Eq. \ref{eq:13b} is
general, and can easily be extended to cover the possibility that the
upward and downward transitions are facilitated by different pulses
whose arrival times coincide.  The theoretical formalism presented is
also well suited explore the use of pulse
shaping\cite{rahav_multidimensional_2010,voronine_manipulating_2007,zhang_selective_2010,roslund_control_2011}
to optimize the Raman signals and highlight desired features.  The
experimental difficulties for x-ray pulse-shaping are daunting, but
technological progress in this area is proceeding very rapidly.

The operator $\alpha_{j}$ is complex and symmetric and therefore
non-Hermitian.\footnote{The effective polarizability is proportional to
the second-order reduced pulse propagator described in
Ref. \protect \onlinecite{biggs_using_2009}} We define its Hermitian
and anti-Hermitian components as
\begin{equation}\begin{split}\label{eq:alphaH}
\alpha^H_j =&(\alpha_j+\alpha^\dagger_j)/2 \\
\alpha^{AH}_j = &(\alpha_j-\alpha^\dagger_j)/2
\end{split} \end{equation}
The matrix elements of these operators are easily found from Eq. \ref{eq:alpha},
\begin{equation}\begin{split}\label{eq:alphaH2}
  \alpha^H_{j;g' g''}&=-\frac{1}{2} \sum_e (\boldsymbol{e_j}\cdot\boldsymbol{V}_{g' e})(\boldsymbol{e_j}\cdot\boldsymbol{V}_{e g''})\mathcal{E}_j^*(\Delta^j_{eg'})\mathcal{E}_j(\Delta^j_{eg''})
   \\ &\times  \textrm{erfi}(-\sigma_j (\Delta^j_{eg'}+\Delta^j_{eg''})/2) \\
  \alpha^{AH}_{j;g' g''}&=-\frac{i}{2} \sum_e (\boldsymbol{e_j}\cdot\boldsymbol{V}_{g' e})(\boldsymbol{e_j}\cdot\boldsymbol{V}_{e g''})\mathcal{E}_j^*(\Delta^j_{eg'})\mathcal{E}_j(\Delta^j_{eg''}).
\end{split}\end{equation}
The 1D-SXRS from Eq. \ref{eq:sxrstime} may be rewritten as
\begin{equation}\label{eq:sxrstime2}
\S_{SXRS}(\tau) = 2\Re \langle \alpha^{AH}_2(\tau)\alpha_1(0)\rangle.
\end{equation}
This signal depends on only on the
anti-Hermitian part of the polarizability for the probe pulse, which
,as Eq. \ref{eq:alphaH2} shows, decreases more rapidly with off-resonant detuning.  We may ignore
vibrational contributions to the elastic component of the signal which
depends on $\alpha_{1;g g}$. The short core-excited state lifetime
precludes a vibrational phase evolving on this potential energy
surface; the x-ray polarizability is diagonal in the vibrational
subspace within the Condon approximation.  Vibrational progressions may appear for long delays as an additional fine structure to the 1D and 2D signals.  These are not included in the present simulations.

\section{Two Dimensional Stimulated Raman: 2D-SXRS}
\label{sec:2dsxrs}
\begin{figure*}[ht]
\centering
  \includegraphics[width=12cm]{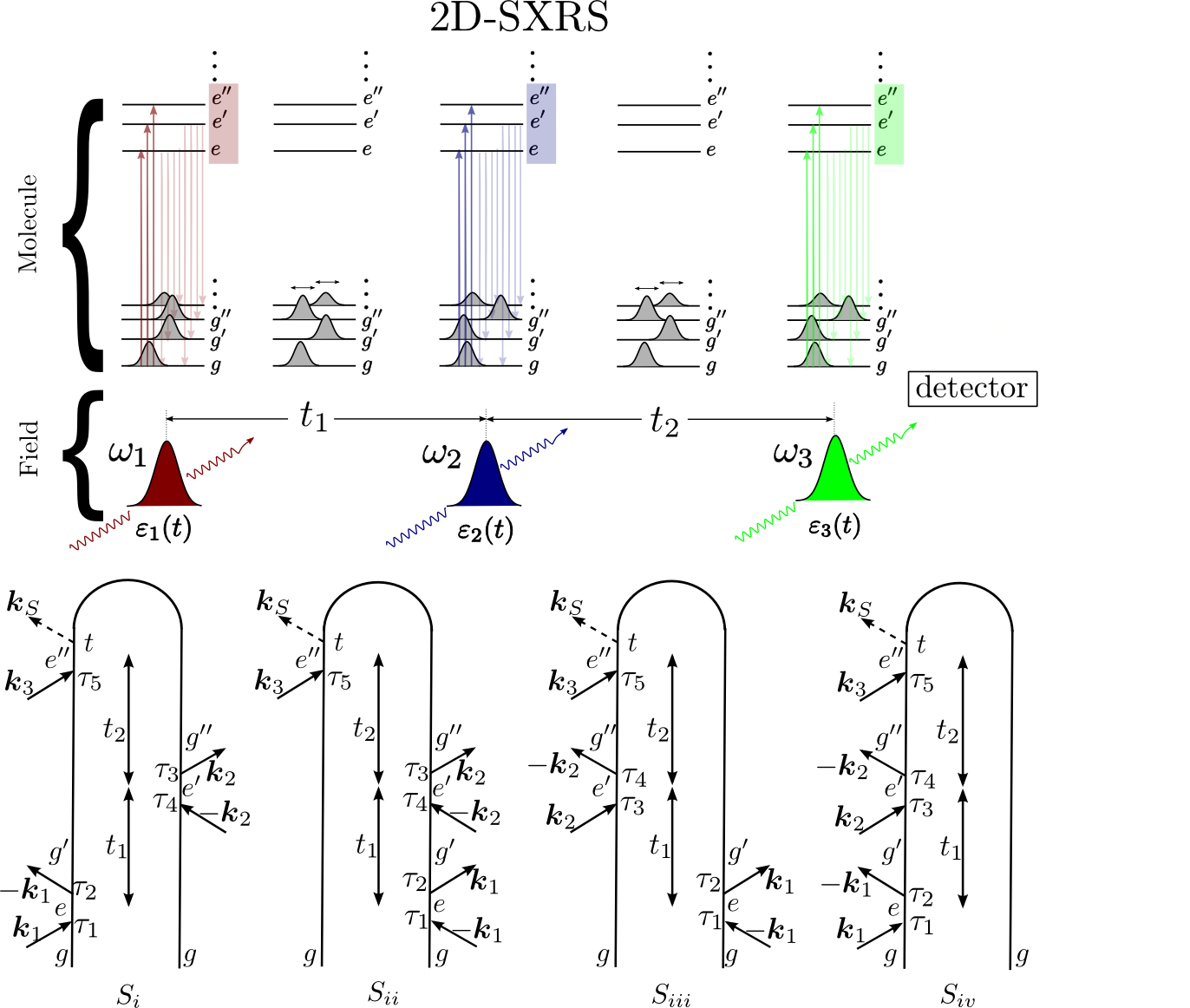}
\caption{Loop diagrams, pulse
  sequence, and energy levels for the 2D-SXRS signal.}
\label{fig:2Dramandiags}
\label{fig:2dsxrstechnique}
\end{figure*}
In 2D-SXRS  the transmitted intensity of
the third pulse depends on two pulse delays. The 2D
signal is related to $\chi^{(5)}$, whereas the 1D-SXRS pump-probe
signal is associated with
$\chi^{(3)}$.\cite{tanimura_two-dimensional_1993} The extra time-delay period
allows more complicated valence electronic dynamics to be prepared and
probed. Again we assume that the delay times $t_1$ and $t_2$ are long
compared to the core-hole lifetime, any
core-excited populations created by the pulses may be safely neglected. The 2D-Raman signal is
represented by the four diagrams in Fig. \ref{fig:2Dramandiags}. Using a
single-particle picture, the first pulse creates an electron-hole pair
in either the ket (diagrams $i$ and $iv$) or the bra (diagrams $ii$
and $iii$).  The second pulse can either create another electron-hole
pair (as in diagrams $i$ and $iii$), or change the electron-hole pair
created by the first (diagrams $ii$ and $iv$). In either case the
pairs created by the first and the second pulses must share either a
common hole or electron in order to survive the trace.  These diagrams
represent different sequences of forward and/or backward time
evolution of valence wavepackets.  These generalize
Fig. \ref{fig:1DRaman} which has a single evolution period.  The
expression for the Raman signal can be read directly from the diagrams
in Fig. \ref{fig:2Dramandiags} and is given by
\begin{equation}\S_{2D-SXRS}(t_1,t_2)=\Im \left[\S_i+\S_{ii}+\S_{iii}+\S_{iv}\right],\end{equation}
where
\begin{equation}\begin{split}\label{eq:2draman1} \S_i &= -\langle\alpha_2^\dagger(t_1) \alpha_3(t_2+t_1)\alpha_1(0)\rangle \\
 \S_{ii}&= \langle\alpha_1^\dagger(0)\alpha_2^\dagger(t_1)\alpha_3(t_2+t_1) \rangle\\
 \S_{iii}&= -\langle\alpha_1^\dagger(0)\alpha_3(t_2+t_1)\alpha_2(t_1)\rangle\\
 \S_{iv} &= \langle\alpha_3(t_2+t_1)\alpha_2(t_1)\alpha_1(0)\rangle \end{split}\end{equation}
\par
This signal may also be recast  as
\begin{equation}\begin{split}\label{eq:2draman1b}
\S_{2D-SXRS}(t_1,t_2) =& 2\Im \big[\langle \alpha^{AH}_3(t_1+t_2){\alpha}_2(t_1){\alpha}_1(0)\rangle \\ & -\langle {\alpha}_1(0) \alpha^{AH}_3(t_1+t_2){\alpha}^\dagger_2(t_1)\rangle \big]
\end{split}\end{equation}
The first term in Eq. \ref{eq:2draman1b}, $\S_{ii}+\S_{iv}$,
is the time-dependent overlap, within a resonance window determined by
the probe pulse, between the ground state and a valence wavepacket
that has interacted with both pump pulses.  The second term,
$\S_{i}+\S_{iii},$ is the overlap between different Raman wavepackets.
\section{Simulations for the oxygen and nitrogen K-edges of NMA}
\label{sec:results}
\subsection{Electronic Structure Calculations}
\begin{figure*}[htbp]
  \includegraphics [width=17cm]{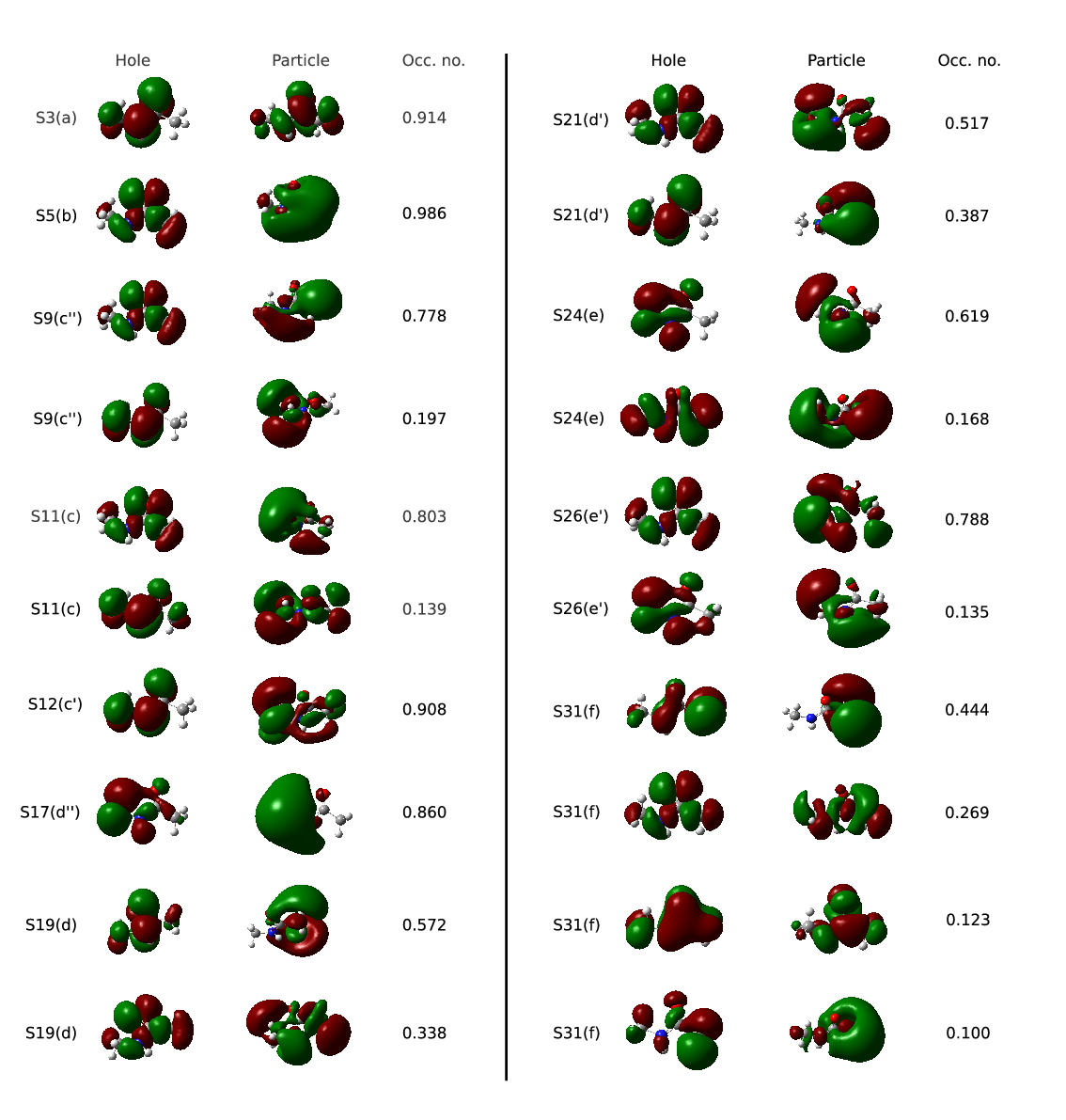}
\caption{ \label{fig:NTO} Natural transition orbitals of the dominant
  excitations in
  Fig. \ref{fig:nma-2DRamanNprobe}-\ref{fig:nma-2DRamanNNN}}
\end{figure*}

The trans-NMA core-excited states are modelled as single
core-hole/virtual orbital electron pairs in the Static Exchange (STEX)
molecular orbital (MO) basis using the orbital approximation. Valence
excited states are treated at the Configuration Interaction Singles
(CIS) level of theory. Both in STEX and CIS Hartree-Fock orbitals are
employed. In order to keep the level of theory consistent we have
calculated the excited states using CIS, which is known to overestimate
the adiabatic excitation energies.\cite{SKF11} More accurate TDDFT
methods to calculate the core excitation energies are under
development.\cite{CysteinePaper} The geometry of \textit{trans}-NMA
was optimized using the quantum chemistry package Gaussian09
\cite{G09} at the B3LYP/6-311G** level. The STEX
model,\cite{HG69,ACVP94,ACVP97} in which a core-ionized Fock matrix is
constructed which includes the core-hole self-consistently, was
implemented in a modified version of the quantum chemistry package
PSI3,\cite{PSI3} with an implementation described in Appendix C of
Ref. \onlinecite{healion_simulation_2011}. All transition frequencies
and dipole moments for core-excited states were calculated using the
orbital approximation at the HF/6-311G** level.  Fig. \ref{fig:NTO}
shows the largest amplitude Natural Transition Orbitals\cite{Martin03}
(NTO) between the ground and excited states which contribute to
Figs. \ref{fig:nma-1DRaman}-\ref{fig:nma-2DRamanNNN}, obtained by a
singular value decomposition of the CIS transition densities.

Using Fermi creation and annihilation
operators for the valence ($c$) and core ($d$) molecular orbitals, the
core-excited states are given by
\begin{equation}
\ket{e}= \ket{jn}=c^\dagger_j d_n \ket{e_o}
\end{equation}
where $\ket{e_o}$ is the lowest energy core excited state. The valence
excited states are
\begin{equation}
\label{eq:gprime} \ket{g'}=\sum_{ai} C^{g'}_{ai} \ket{ai}=
\sum_{ai}  C^{g'}_{ai}  c^\dagger_a c_i \ket{g},
\end{equation}
where $C^{g'}_{ai}$ are the CI coefficients for state $\ket{g'}$.

\begin{table}[htbp]
\caption{\label{tab:1}CIS valence excited state energies $\{
  \omega_{g'g}\}$.  }
\begin{center}\begin{tabular}{l r c l r } \hline \hline
State & Energy (eV) & & State& Energy (eV) \\ \hline
 $S_1$ & 6.92 & & $S_{26}$ & 13.64 \\
 $S_2$ & 8.14 & & $S_{27}$ & 13.80 \\
 $S_3$ & 8.95 & & $S_{28}$ & 13.95 \\
 $S_4$ & 9.99 & & $S_{29}$ & 14.04 \\
 $S_5$ & 10.23 & & $S_{30}$ & 14.14 \\
 $S_6$ & 10.92 & & $S_{31}$ & 14.25 \\
 $S_7$ & 11.32 & & $S_{32}$ & 14.27 \\
 $S_8$ & 11.35 & & $S_{33}$ & 14.32 \\
 $S_9$ & 11.41 & & $S_{34}$ & 14.59 \\
 $S_{10}$ & 11.66 & & $S_{35}$ & 14.59 \\
 $S_{11}$ & 11.78 & & $S_{36}$ & 14.85 \\
 $S_{12}$ & 11.83 & & $S_{37}$ & 14.89 \\
 $S_{13}$ & 11.86 & & $S_{38}$ & 14.91 \\
 $S_{14}$ & 12.11 & & $S_{39}$ & 14.91 \\
 $S_{15}$ & 12.30 & & $S_{40}$ & 15.09 \\
 $S_{16}$ & 12.35 & & $S_{41}$ & 15.10 \\
 $S_{17}$ & 12.54 & & $S_{42}$ & 15.22 \\
 $S_{18}$ & 12.64 & & $S_{43}$ & 15.25 \\
 $S_{19}$ & 12.68 & & $S_{44}$ & 15.39 \\
 $S_{20}$ & 12.71 & & $S_{45}$ & 15.50 \\
 $S_{21}$ & 12.84 & & $S_{46}$ & 15.59 \\
 $S_{22}$ & 13.01 & & $S_{47}$ & 15.63 \\
 $S_{23}$ & 13.44 & & $S_{48}$ & 15.69 \\
 $S_{24}$ & 13.49 & & $S_{49}$ & 15.81 \\
 $S_{25}$ & 13.55 & & $S_{50}$ & 15.92	\\ \hline \hline
 \end{tabular}  \end{center} \end{table}

\subsection{Valence Excitations and UV Absorption}
Earlier simulations on trans-NMA presented in
Ref. \onlinecite{healion_simulation_2011} approximated the valence
excited states as single electron-hole pairs in the HF MO basis.  The
lowest energy valence-excited state then involved a transition from
the highest occupied orbital to the lowest virtual orbital, a HOMO to
LUMO transition.  The excitation energy was approximated as the
difference in MO energies, which gave 14.35 eV. Here we use a higher
level treatment of valence excited states obtained by diagonalizing
the singly-excited block of the Hamiltonian. The lowest-energy
excitation is now found to be 6.92 eV (see Table \ref{tab:1} for a
list of the CIS energies), which is
consistent with experimental vacuum-UV absorption results from
trans-NMA.\cite{kaya_vacuum_1967} The valence excitations and the
Raman peaks presented here (between 7 and 18 eV) are very different
from those in Ref. \onlinecite{healion_simulation_2011}.

\begin{figure}[htbp]
  \includegraphics[width = 7cm] {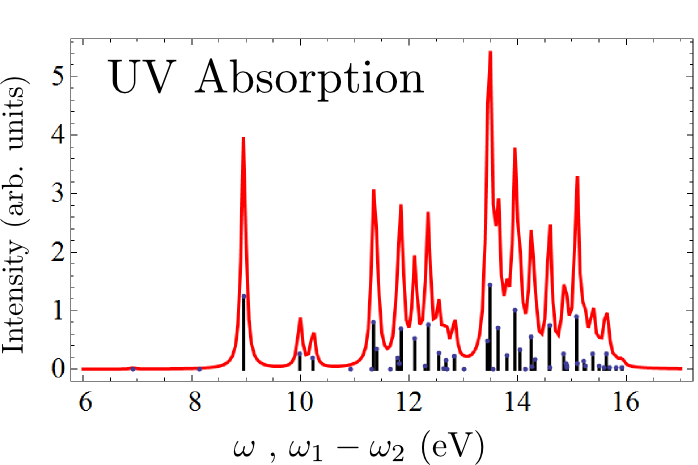}
\caption{\label{fig:nma-UV} Calculated UV-absorption spectrum of NMA.}
\end{figure}

The calculated UV absorption spectrum is displayed in Fig. \ref{fig:nma-UV}.  The molecule is taken to be initially in the
ground state and the $\delta(\omega-\omega_{g'g})$ in Fermi's golden
rule was replaced by a Lorentzian, with an effective lifetime
$\Gamma_{g'},$ to account for the linewidth of valence electronic
transitions
\begin{equation}
\S_{UV}(\omega) = \sum_{g'} \frac{\left| \mu_{g'g} \right|^2 \Gamma_{g'}}{(\omega-\omega_{g'g})^2+\Gamma_{g'}^2}.
\end{equation}
An accurate estimate of $\Gamma_{g'}$ would require
potential energy surface calculations and characterization of any
conical intersections or dissociative states which could cause decay
of the excited state population in $\vert g' \rangle$. We
set $\Gamma_{g'}=0.05$ eV for all valence excitations in all
simulations for display purposes.

\subsection{Frequency Domain XANES and RIXS Spectra}
\begin{figure}[htbp]
  \includegraphics[width = 8cm] {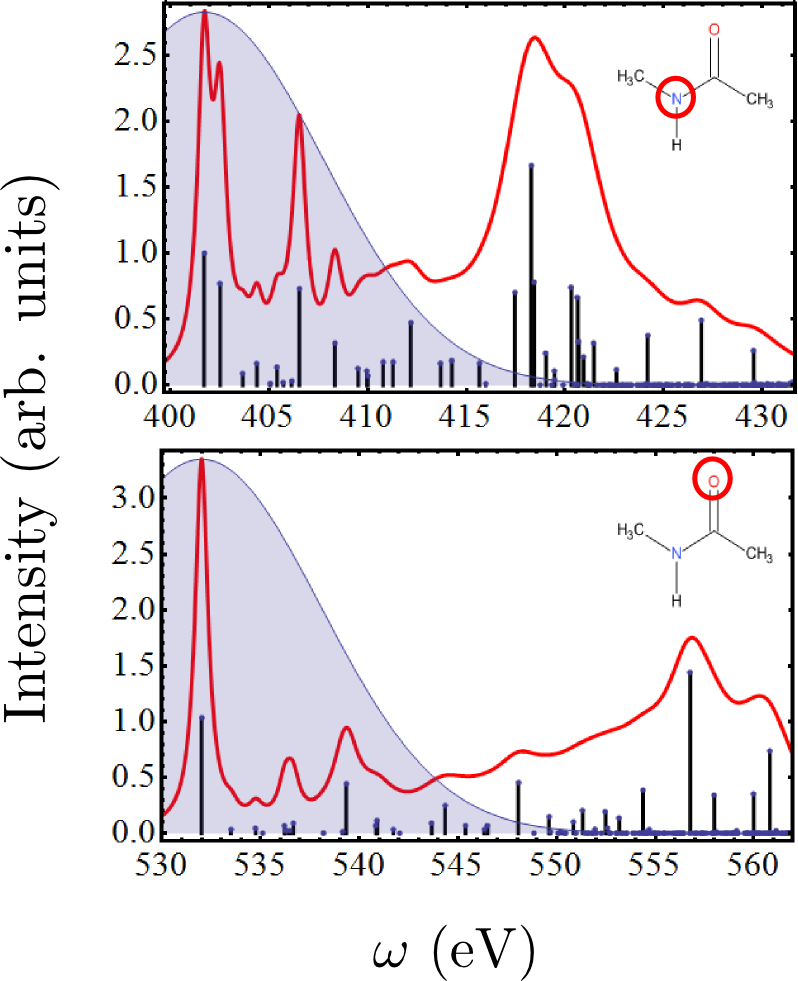}
\caption{\label{fig:nma-xanes-isotropic}Simulated XANES from trans-NMA at the nitrogen (top) and oxygen (bottom)
K-edge.  The stick spectra (black lines) have been convoluted with a lineshape function (see text) to give the
spectra in red.  Shown in blue are the power spectra for the Gaussian pulses used in the time-domain experiments described here.}
\end{figure}
In Fig. \ref{fig:nma-xanes-isotropic} we present the x-ray Absorption
Near-Edge Sstructure (XANES) spectra calculated for an isotropic
trans-NMA sample, excited at the nitrogen and oxygen K-edges. The
XANES and the UV absorption spectra were calculated using a Lorentzian lineshape to account for core lifetime
broadening and pure dephasing
\begin{equation}
\S_{XANES}(\omega) = \sum_e \frac{\left|\mathbf{\mu}_{eg} \right|^2 \Gamma_{e}}{(\omega-\omega_{e g})^2+\Gamma_{e}^2}.
\end{equation}
As in Ref. \onlinecite{healion_simulation_2011}, we set the first
XANES transition for nitrogen to 401.7 eV and to 532.0 eV for oxygen,
to match experiment.\cite{bearden_reevaluation_1967}
Each XANES peak represents an excitation from the core-orbital to a
bound virtual orbital.  Orbitals with energies above the ionization
potential are coupled to a continuum of photoelectron states. The
linewidth for nitrogen was taken as: $\Gamma_{eN}$ = 0.4 eV for
energies up to 408 eV, and ramped up to 1.5 eV at 415 eV and held
constant for higher energies.  A similar form was followed for
oxygen: $\Gamma_{eO}$ = 0.4 eV for energies up to 537 eV, and ramped
up to 1.5 eV at 544 eV and held constant thereafter. These
phenomenological linewidths, which were used in earlier studies to
match experimental XANES spectra of small nitrogen and oxygen
containing organic
molecules,\cite{schweigert_probing_2007,schweigert_double-quantum-coherence_2008}
can reflect a variety of broadening mechanisms, including vibrational
motion of the core-ionized molecule, autoionization of the excited
electron and direct photoionization of the core into a photoelectron
state. An exact treatment of these effects would require a
Wigner-Weisskopf treatment of the decay of the bound excited electron
into a continuum of photoelectron states, and potential energy surface
calculations of these core-excited states. We use the atomic
core-lifetime given in Section \ref{sec:1dsxrs} in the calculation of
$\alpha$.

To set the stage we first calculate the traditional frequency-domain spontaneous Raman (RIXS) spectra. These should
serve as a reference for comparing with the 1D-SXRS signals.  The optical resonant Raman
technique probes which vibrational modes of a molecule are perturbed by a given
electronic excitation.  Only those modes whose potential is different
in the ground and excited electronic states have
Franck-Condon activity.  In RIXS a core electron is excited into an
empty virtual orbital and then de-excited.  Inelastic losses
representing valence excitations can be created directly, when an
electron other than the excited core electron drops into the
core-hole, or indirectly when the Coulomb potential of the
transiently-created core-hole creates valence
excitations. \cite{brink_correlation_2006,ament_resonant_2011} The
energy satellite spectra for the indirect process represent valence
shake-up states, and are dependent on the excitation frequency.

In RIXS a monochromatic x-ray beam ($\omega_1$) impinges on the
molecule, and the scattered radiation ($\omega_2$) is frequency
resolved, with peaks recorded vs. $\omega_1-\omega_2=\omega_{g'g}$.
Here we take the incident beam $\omega_1$ to be polarized, and the
scattered light $\omega_2$ is also sent through a polarization filter
before detection (detected parallel to the excitation polarization).
The Kramers-Heisenberg formula  with a Lorentzian,
linewidth $\Gamma_{g'}=$0.05 eV, gives
\begin{equation}\label{eq:rixs}
\S_{RIXS}(\omega_1,\omega_2)= \sum_{g'} \frac{\left| \tilde{\alpha}_{g'g}(\omega_1) \right|^2 \Gamma_{g'}}{(\omega_1-\omega_2-\omega_{g'g})^2 +\Gamma_{g'}^2},
\end{equation}
where
\begin{equation}\label{eq:CWalpha}
\tilde{\alpha}_{g'g}(\omega_1) = \sum_e \frac{(\mathbf{e}_I \cdot \mathbf{\mu}_{g'e})(\mathbf{e}_I \cdot \mathbf{\mu}_{eg})}{\omega_1-\omega_{eg}+i \Gamma_e}.
\end{equation}
is the frequency-dependent resonant polarizability.
Eq. \ref{eq:CWalpha} applies to an oriented molecule in the lab frame.  To
account for the isotropic distribution of molecules,
we orientationally averaged Eqs. \ref{eq:rixs} and \ref{eq:CWalpha}
assuming parallel excitation and detection
polarization.\cite{gelmukhanov_resonant_1994}

\begin{figure}[htbp]
  \includegraphics[width = 8.5cm] {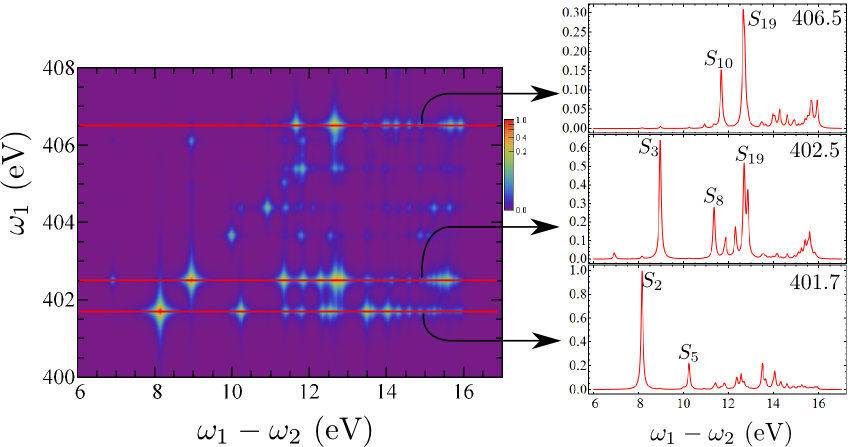}
\caption{\label{fig:nma-rixs-Nedge}Calculated RIXS signal at the
  nitrogen K-edge from trans-NMA.}
\end{figure}
\begin{figure}[htbp]
  \includegraphics[width = 8.5cm] {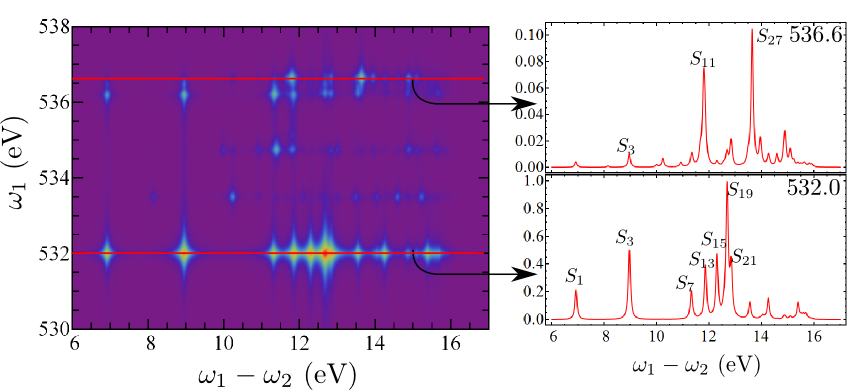}
\caption{\label{fig:nma-rixs-Oedge}Calculated RIXS signal from trans-NMA at the
  oxygen K-edge.  }
\end{figure}

The calculated RIXS spectra with the incident light tuned to the
nitrogen and oxygen K-edge are displayed in
Figs. \ref{fig:nma-rixs-Nedge} and \ref{fig:nma-rixs-Oedge},
respectively. The RIXS intensity falls off as the laser frequency is
detuned from the strong core-edge transition, as expected from
Eq. \ref{eq:CWalpha}.  The peak pattern changes with
$\omega_1$, since the intensity also depends on the transition dipole
between the valence-excited state in question and the core-excited
state in resonance with the laser frequency.

When $\omega_1=401.7$ eV, the transition frequency for the first N1s
core-excited state, the valence-excited states $S_2$ and $S_5$ are
strongly Raman active.  Tuning $\omega_1$ to be resonant with the second
core-excited state causes these peaks to be replaced by a strong $S_3$
peak.  Changing the excitation frequency to 406.5 eV (the ninth
core-excited state) we see the absence of $S_2,$ $S_3,$ and $S_5$
peaks, and the dominant peak is now from the $S_{19}$ state.  The
oxygen RIXS spectrum, shown in Fig. \ref{fig:nma-rixs-Oedge}, shows a
similar pattern.  The signal is strong at the core edge, $\omega_1 =
532.0$ eV, and is dominated by the $S_3$, $S_{15},$ and $S_{19}$
peaks.  Increasing the excitation frequency to 536.6 eV gives strong
peaks for $S_{11}$ and $S_{27}$. This atomic selectivity makes RIXS a
powerful tool for probing the delocalization of valence excited
states.\cite{glatzel_electronic_2004,groot_core_2008}

While the valence-excited states that contribute to the UV absorption
spectrum and the RIXS and SXRS spectra are the same, their
intensities are radically different, as can be seen by comparing
Fig. \ref{fig:nma-UV} with Figs. \ref{fig:nma-rixs-Nedge} and \ref{fig:nma-rixs-Oedge}.  This is because
the absorption is determined by the dipole moment between the HF ground state
and the valence-excited state, while the Raman spectra depend on the
polarizability resonant with a given core-excited state.

\subsection{Time-domain Stimulated X-Ray Raman Spectroscopy: 1D-SXRS}
In both 1D-SXRS and RIXS\cite{dhar_time-resolved_1994}, information on
valence excited states is obtained through transient
excitation of core excited states. 1D-SXRS has two advantages over its
frequency domain counterpart. First, because the interaction times are
constrained by the field pulse envelopes, it should be possible to
probe electronic dynamics directly by preceding the two Raman pulses
with an initiation pulse.  This pulse could prepare the molecule in
some nonstationary state which may then be probed by an SXRS
experiment, in much the same way as is currently done in vibrational
spectroscopy.\cite{fang_mapping_2009,kukura_femtosecond_2007} We do
not pursue this course here, rather we look at another advantage
inherent in the time-domain experiment.  We can use a
two color scheme, where the two pulses are tuned to be resonant with
different core transitions, providing an additional experimental knob to turn.
The differences between SXRS spectra with different pulse
configurations could help shed light on the nature of the valence
excited states of a molecule.  The two diagrams in
Fig. \ref{fig:1dsxrstechnique} recast the SXRS spectrum as a two-slit
experiment where valence wave packets created on either the ket or the
bra by the first pulse are probed by another wave packet created in
the ket by the second pulse.

The calculated 1D-SXRS signal is displayed as the Fourier transform of
Eq. \ref{eq:sxrstime},
\begin{widetext}
\begin{equation}\begin{split}\label{eq:sxrsfreq}
\S_{SXRS}(\Omega)  =& \int_{0}^\infty d\tau \S_{SXRS}(\tau) \exp \left[i \Omega \tau \right] \\
=& -\sum_{g'}\frac{\Re\left(\alpha _{2;g g'}\alpha _{1;g'g}+(\alpha _{1;g g'})^*\alpha _{2;g'g}\right) \left(\Gamma _{g'}-i \Omega \right)+\Im\left(\alpha _{2;g g'}\alpha _{1;g'g}-(\alpha _{1;g g'})^*\alpha _{2;g'g}\right) \omega _{g'g}}{ \left(\Gamma _{g'}^2-2 i \Gamma _{g'} \Omega -\Omega ^2+\omega _{g'g}^2\right)}
\end{split}\end{equation} \end{widetext}
Unlike RIXS (Eq. \ref{eq:rixs}), $\S_{SXRS}(\Omega)$ results from the
interference of the two pathways in Fig. \ref{fig:1DRaman} and
may not be written as an amplitude squared.  The signal is collected in
the time domain, and Fourier transformed numerically.  It is important
that the interpulse delay be longer than the core-hole lifetime ($>10$
fs). In this case, any interaction with the pulses that results in a
core-excited population will decay to an ionized state through an
Auger process, with transitions that are detuned from the resonant
excitation with the probing pulse.  If the core-hole lifetimes were
long compared to the delay, the signal would be distorted by an
interference between the wavepacket produced on the core-excited
states and the ground valence electronic
wavepacket.\cite{schweigert_probing_2007} We also neglect a
$\tau$-independent contribution to the signal, resulting from elastic scattering with $g'=g$.
This can be removed experimentally.

\begin{figure}[htbp]
  \includegraphics[width = 8.5cm] {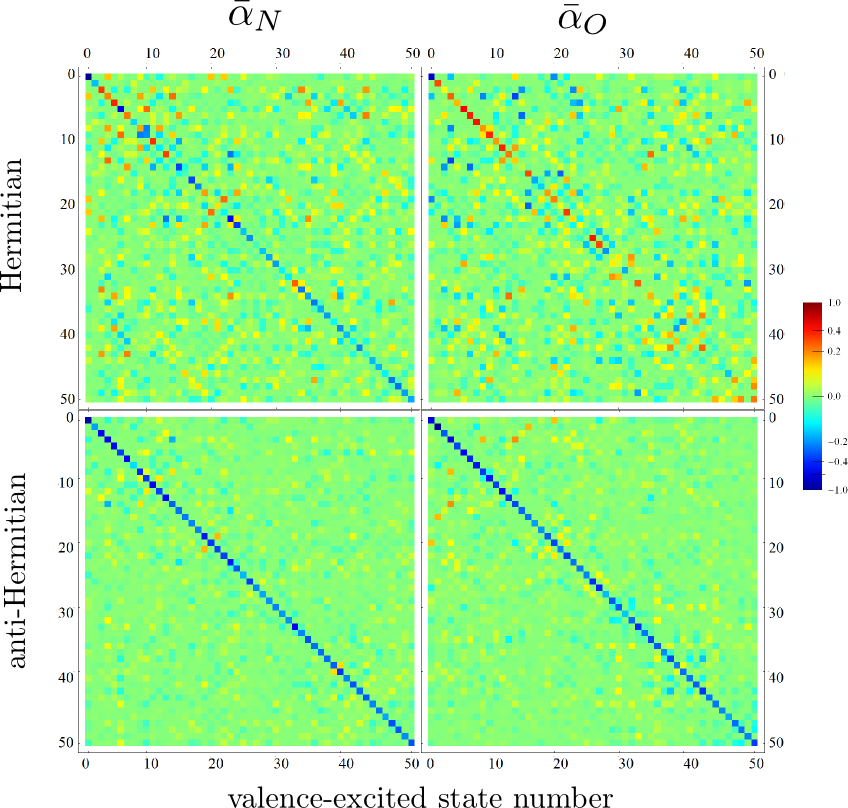}
\caption{\label{fig:nma-2Dalphas}Hermitian and anti-Hermitian parts of the  effective isotropic
  polarizabilities (Eq. \ref{eq:alpha}) for the two pulses used in
  our simulations corresponding to the nitrogen and oxygen K-edge excitations, plotted using an arcsinh nonlinear scale (shown on the right). The Hermitian part is purely real, while the anti-Hermitian part
  is purely imaginary.  The axes are labeled by the state numbers, 0 for the
  ground state, 1 for $S_1$, etc. State assignments can be found in
  Table \ref{tab:1}. }
 \end{figure}
The Hermitian and anti-Hermitian parts of the resonant polarizability
matrix $\alpha$ averaged over the pulse bandwidth are shown in
Fig. \ref{fig:nma-2Dalphas}.  We assume pulses with Gaussian
envelopes, durations of $\sigma_j = 77 \, \mathrm{as}$ ( $1/\sigma_j
\approx 8.5 \, \mathrm{eV}),$ and center frequencies set to the N and
O K-edge transitions (401.7 eV and 532.0 eV, respectively). The axes
are labeled by the valence excited state number, with 0 referring to
the ground state.  We display the isotropic polarizability
$\bar{\alpha}_j$, obtained by replacing the direction cosine term
$(\boldsymbol{e_j}\cdot\boldsymbol{V}_{g'
  e})(\boldsymbol{e_j}\cdot\boldsymbol{V}_{e g''})$ with
$\boldsymbol{V}_{g' e}\cdot\boldsymbol{V}_{eg''}$ in Eq. \ref{eq:13}.
Otherwise, it would be necessary to adopt a given molecular lab-frame
orientation to visualize $\alpha_j$.  This 1D SXRS spectrum probes
matrix elements of the effective polarizability in Eq. \ref{eq:alpha}
between the ground state and the set of singly excited state,
$\alpha_{j;g g'}$ for $j=\mathrm{N},\mathrm{O}$.  The 1D-SXRS signal
is a function of the top row of this symmetric matrix.

\begin{figure*}[htbp]
  \includegraphics[width = 14.4cm] {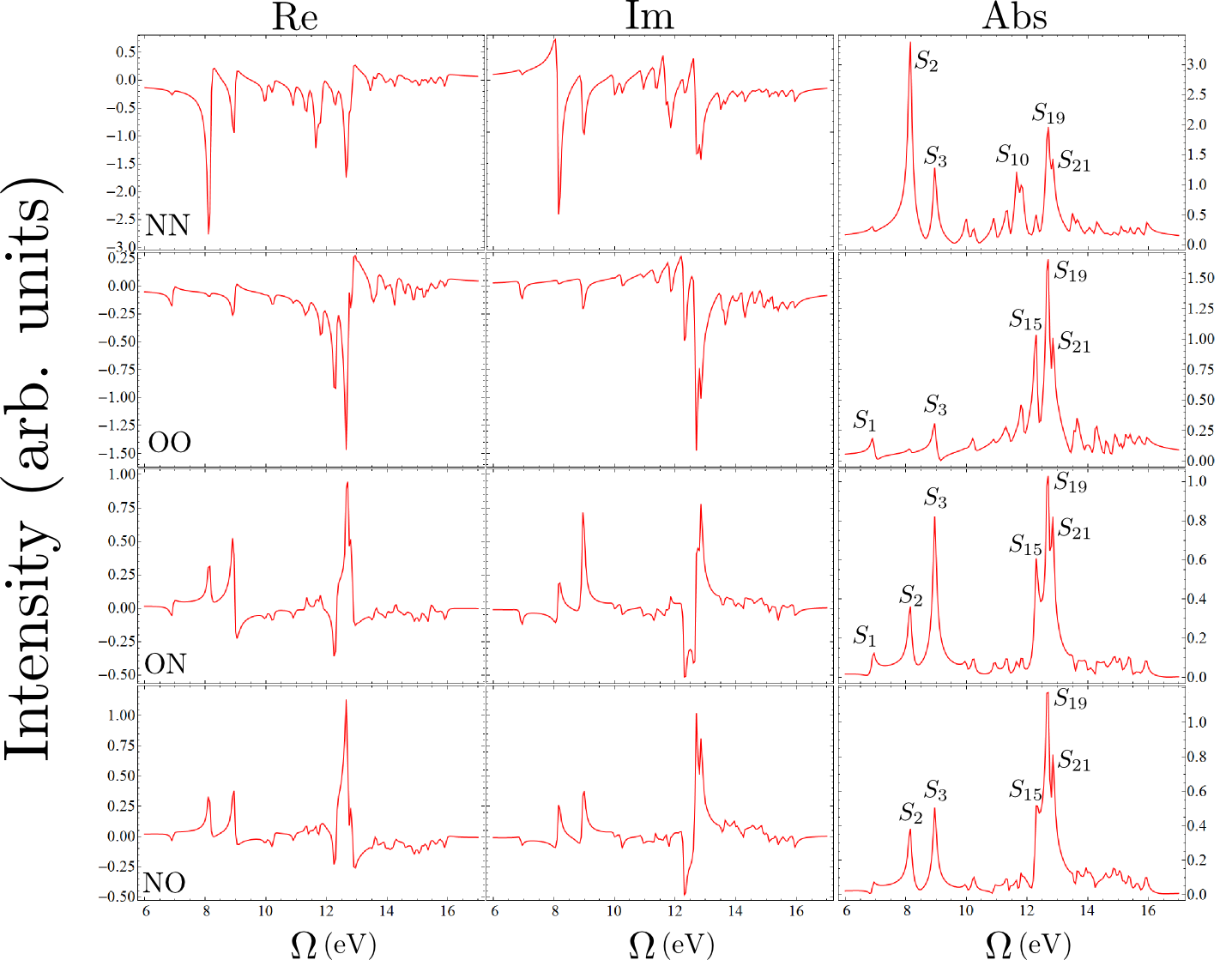}
\caption{\label{fig:nma-1DRaman}Calculated SXRS spectra from
  trans-NMA, in which both pulses are polarized parallel to the lab
  frame V axis.  The pulses are Gaussian, 181 as FWHM in intensity,
  with center frequency set to either 401.7 eV (N) or 532.0 (O). From
  left to right we show the real part, imaginary part, and modulus of
  Eq. \ref{eq:sxrsfreq}.  The two-color signals (bottom two rows) have
  their pulse sequences given from left to right in chronological
  order, i.e. the ON signal results from having the O pulse come first
  and the N pulse come second. }
\end{figure*}
The simulated 1D-SXRS signals are shown in Fig. \ref{fig:nma-1DRaman}.
Parallel field polarizations are used, and the signal
is averaged over an isotropic distribution of
molecules.\cite{andrews_three-dimensional_1977} The 4 rows represent
the four possible pulse configurations, e.g. row 3 is labeled ON
signifying that the first pulse is tuned to the oxygen and the second
pulse to the nitrogen K-edge.  The three columns show the real,
imaginary, and modulus of Eq. \ref{eq:sxrsfreq}.

The one-color, time-domain signals in the top two rows of
Fig. \ref{fig:nma-1DRaman} resemble their frequency-domain
counterparts of Figs. \ref{fig:nma-rixs-Nedge} and
\ref{fig:nma-rixs-Oedge}.  In the N-edge RIXS spectrum, the $S_2$ and
$S_3$ peaks are strong when the excitation frequency is set to 401.7
eV and 402.5 eV, respectively.  Both peaks show up prominently in the
NN 1D-SXRS spectrum.  Since the excitation must be broadband to coherently excite valence excited states,
1D-SXRS does not select a given $e$ state.  An interesting possibility will be to use a combination of narrow band excitation
combined with broadband stimulated de-excitation, as is commonly done
in optical Raman spectroscopy,\cite{kukura_femtosecond_2007} to regain
this selectivity over core-excited states.

Two-color 1D-SXRS signals have no RIXS analogue.  In the third and
fourth rows of Fig. \ref{fig:nma-1DRaman} we show the ON and NO
spectra.  The moduli of these two signals are virtually identical but
differ in their real and imaginary parts.  The $S_2$ peak is mostly
absent from the OO signal, very strong in the NN signal, and moderate
in the ON and NO signals.  However, the two-color signals are not
simply averages of the one-color signals from the top two rows.  For
instance, the $S_{10}$ peak that shows up prominently in the NN
spectrum is missing completely from the two-color spectra.  Thus,
two-color 1D-SXRS provides a different window into the interaction
between core-hole excitation and valence-excited states.

\subsection{Two-dimensional stimulated Raman: 2D-SXRS}
The 2D-SXRS signal in Eq. \ref{eq:2draman1} is the product of
$\alpha_{1;g' g}$ and $\alpha_{k;g g''},$ from the top row of the
polarizability matrix with a third term, $\alpha_{l;g' g''}$ from its
interior (see Fig. \ref{fig:nma-2Dalphas}).  The 1D-SXRS signal
(Eq. \ref{eq:sxrsfreq}) only depends on the first row and column
of the polarizabilities; 2D-SXRS thus carries information about
correlations between dynamics in the two time periods not available in
the 1D signals.

The 2D signals will be displayed in the frequency domain
\begin{equation}\begin{split}\label{eq:2dsxrs}
\S_{2D-SXRS}(\Omega_1,\Omega_2)=& \int_{0}^\infty dt_1 \\ & \times \int_{0}^\infty dt_2  e^{i \Omega_1 t_1+i \Omega_2 t_2}\S_{2D-SXRS}(t_1,t_2),
\end{split}\end{equation}
The contribution to the 2D signal coming from, for example, diagram
$\S_i$ (see Eq. \ref{eq:2draman1}) is given by
\begin{widetext}
\begin{equation}\begin{split}\label{eq:2dsxrsA}\frac{\Re\left(\left(\alpha _{2;g g''}\right){}^*\alpha _{3;g''g'}\alpha _{1;g'g}\right) \left(\Gamma _{g'}{}^2-\omega _{g'g}\omega _{g'g''}-\Omega _1 \Omega _2-i \Gamma _{g'} \left(\Omega _1 +\Omega _2\right)\right)}{ \left(\omega _{g'g}^2+\left(\Gamma _{g'}-i \Omega _1\right){}^2\right) \left(\omega _{g'g''}^2+\left(\Gamma _{g'}-i \Omega _2\right){}^2\right)} \\
+\frac{\Im\left(\left(\alpha _{2;g g''}\right){}^*\alpha _{3;g''g'}\alpha _{1;g'g}\right)\left(\Gamma _{g'} \left(\omega _{g'g}+\omega _{g'g''}\right)-i \left(\omega _{g'g''} \Omega _1+\omega _{g'g} \Omega _2\right)\right)}{ \left(\omega _{g'g}^2+\left(\Gamma _{g'}-i \Omega _1\right){}^2\right) \left(\omega _{g'g''}^2+\left(\Gamma _{g'}-i \Omega _2\right){}^2\right)}
\end{split}\end{equation}
\end{widetext}
and the other contributions follow similarly.  As in the 1D signal, we
remove any zero frequency contributions (for $\Omega_1$ and
$\Omega_2$) prior to plotting.

Fig. \ref{fig:nma-2Dalphas} shows that the anti-Hermitian part of
$\alpha,$ for both pulses is mostly diagonal, while the Hermitian part
has significant off-diagonal contributions.  Since the 2D signal
depends only on the anti-Hermitian part of the polarizability for the
probe pulse, the probe acts as a filter through which the various
valence-excited wavepackets created by the pump pulses interfere with
each other.

\begin{figure}[htbp]
  \includegraphics[width = 8.5cm] {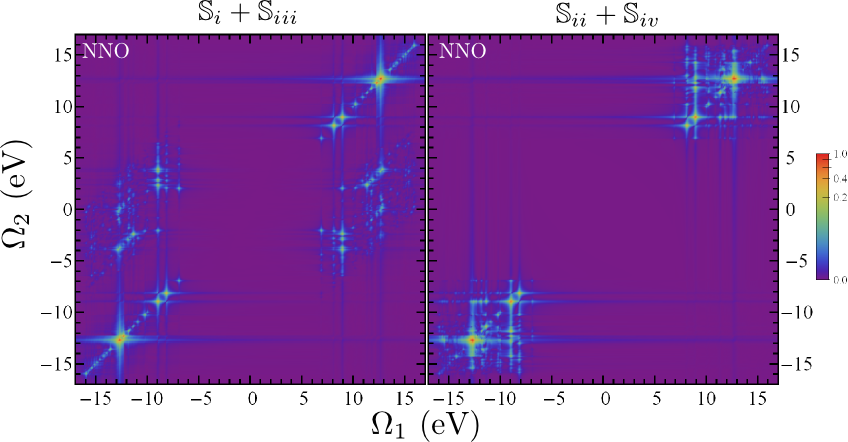}
\caption{\label{fig:nma-2DNNOdiags}Simulated 2D-XRS spectrum from
  trans-NMA using an NNO pulse configuration, plotted as the modulus
  of the Fourier transform and separated into the contributions from
  the two types of diagrams in Fig. \ref{fig:2Dramandiags}.  The
  labels refer to the pulse center frequency and polarization of the
  three pulses ordered chronologically from left to right.  In the NNO
  signal, the first and second pulses have their center frequency
  resonant with the nitrogen K-edge transition, and the third pulse is
  likewise tuned to the oxygen K-edge. Signals are plotted using an
  arcsinh nonlinear scale (see color bar) to highlight weak features. }
\end{figure}
In Fig. \ref{fig:nma-2DNNOdiags} the modulus of Eq. \ref{eq:2dsxrs} is
shown for the NNO pulse sequence, in which the first two pulses
are resonant with nitrogen and the last with oxygen.  Diagonal peaks are the
largest features due to the large diagonal matrix elements of
$\alpha^j$.  The contributions $\S_i$ and $\S_{iii}$ contain diagonal
peaks for $g=g''$.  Likewise, in $\S_{ii}$ and $\S_{iv}$, we get a
diagonal peak when $g'=g''$. When different valence excited states are
involved, off-diagonal peaks of two different varieties are found.
$\S_{ii}$ and $\S_{iv}$ contain peaks at
$(\Omega_1,\Omega_2)=(\omega_{g g'},\omega_{g g''})$, i.e. $\omega_2$
is a valence excitation frequency of the system.  In the other
diagrams, $\Omega_{2}=\omega_{g'g''}$ is the difference between system
excitation frequencies.  Inelastic peaks from the $\S_{i}$ and
$\S_{iii}$ contributions are therefore spectrally removed from the
much stronger elastic peaks.  Since each diagram contributing to the
signal defined in Eq. \ref{eq:2dsxrs} has inversion symmetry, further
plots will omit the portion of the signal where $\Omega_1<0$.

Off-diagonal peaks are weaker than diagonal peaks because the diagonal
elements of $\alpha_j$ are much larger than the off-diagonal
elements. To enhance these features, we plot the 2D signals using the
nonlinear scale,
\begin{equation}\textrm{arcsinh}(\S) = \textrm{ln}\left( \S+\sqrt{1+\S^2}\right).
\end{equation}
This scaling function interpolates between linear (for small \S) and logarithmic (for large \S)
scaling and shows both weak and strong features (the modulus of the
signal is taken prior to applying this scaling function).

\begin{figure*}[htbp]
  \includegraphics[width = 14.4cm] {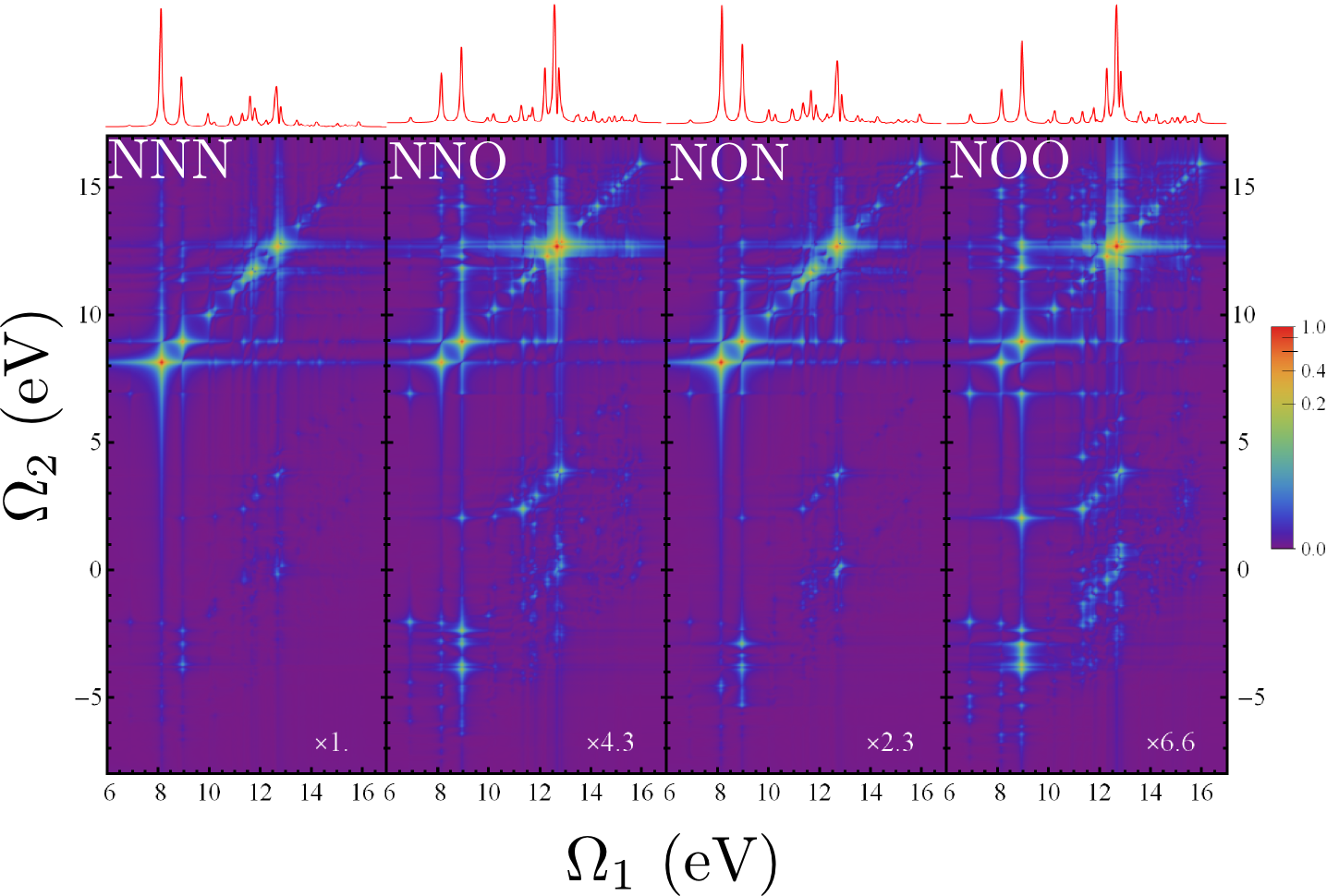}
\caption{\label{fig:nma-2DRamanNprobe}Simulated 2D-XRS spectra from trans-NMA, plotted as the modulus of the Fourier transform.  The labels refer to the pulse center
  frequency and polarization of the three pulses ordered
  chronologically from left to right.  In the NNO signal, the first
  and second pulses have their center frequency resonant with the nitrogen K-edge transition, and
  the third pulse is likewise tuned to the oxygen K-edge. Signals are plotted using an arcsinh nonlinear scale to highlight weak features.
  Traces of each signal along the diagonal are shown in red on top of each signal.}
\end{figure*}
\begin{figure*}[htbp]
  \includegraphics[width = 14.4cm] {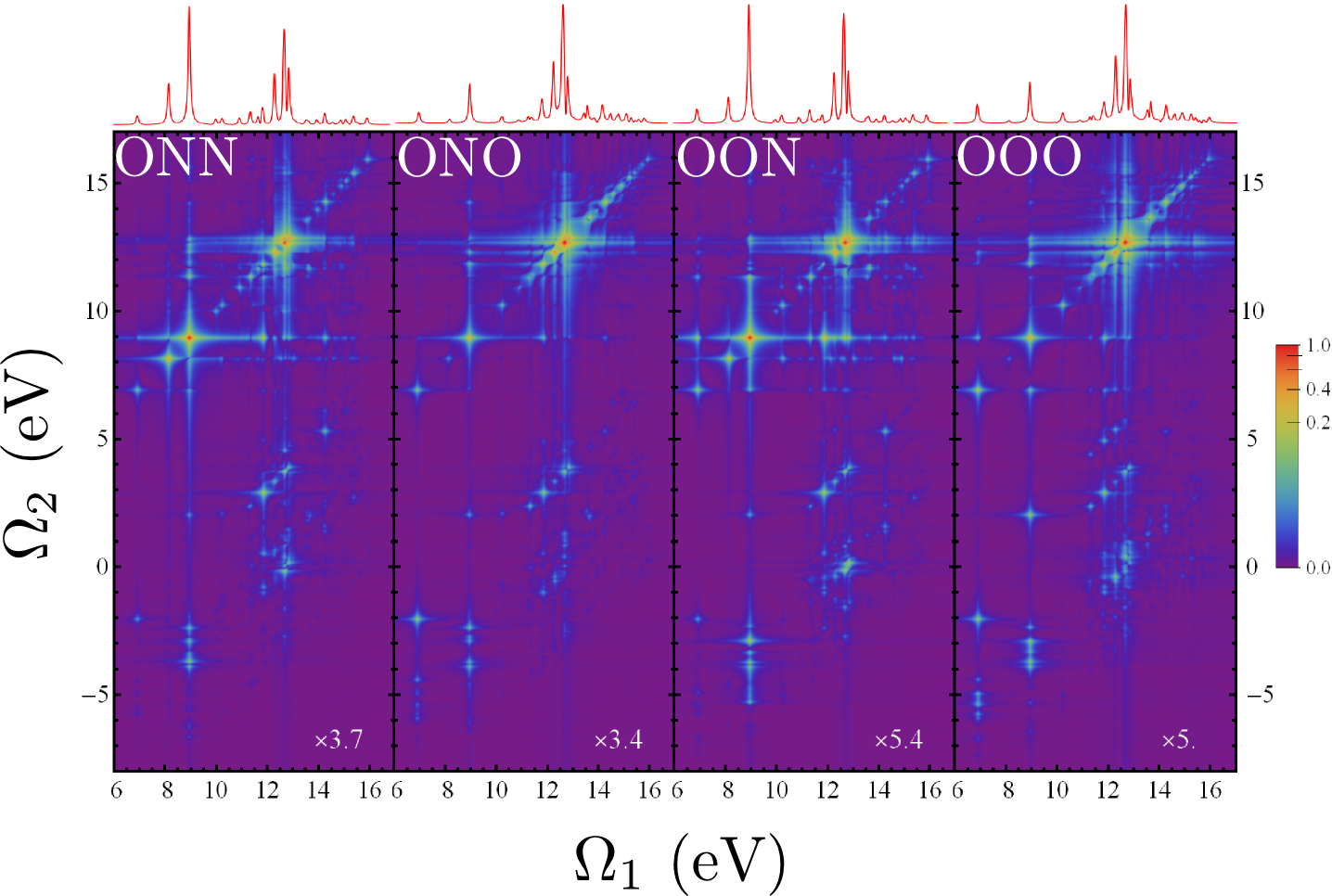}
\caption{\label{fig:nma-2DRamanOprobe}Same as
  Fig. \ref{fig:nma-2DRamanNprobe} for the other pulse
  configurations.}
\end{figure*}

Plots of all 8 two-color 2D-SXRS signals are given in Figs. \ref{fig:nma-2DRamanNprobe} and \ref{fig:nma-2DRamanOprobe}.
Detailed interpretation of this information, relating the intensities
of the peaks in a 2D spectrum to the shape of the various orbitals and their underlying many-body wavefunctions
will require further study.  These signals may also prove to be a valuable test for the quality of electronic structure
calculations on molecular systems.
\par
The 1D spectra with the NO and ON pulse configurations are very
similar, as may be seen in the bottom two panels of the right column of
Fig. \ref{fig:nma-1DRaman}.  However, the 2D signals, especially in
the spectral region $6 \, \mathrm{eV} >\omega_2>-6 \, \mathrm{eV}$,
show a much greater sensitivity to pulse combinations. Note the
radically different spectrum that results in
Fig. \ref{fig:nma-2DRamanNprobe} when the last two pulses are
interchanged in going from the NNO to the NON configuration.

\begin{figure}[htbp]
  \includegraphics[width = 8.5cm] {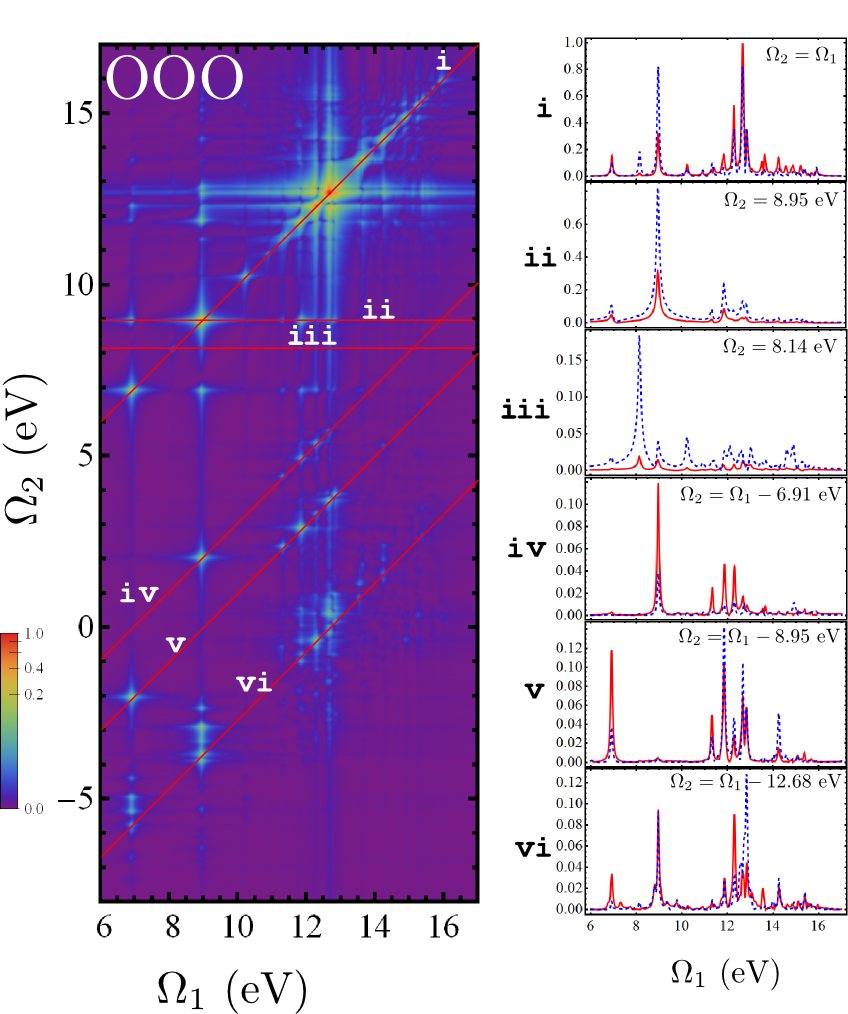}
\caption{\label{fig:nma-2DRamanNNN}(Left) An enlarged version of the  OOO spectrum from Fig. \ref{fig:nma-2DRamanNprobe}, plotted using a nonlinear scale shown on the color bar to the left.
(Right) Horizontal and diagonal slices, plotted using a linear scale, of the 2D spectrum on the left (in red) plotted
together with the corresponding traces from the corresponding OON (dashed, blue) to highlight the effect of changing the probe pulse in the three-pulse sequence.}
\end{figure}

In Fig. \ref{fig:nma-2DRamanNNN} we show the OOO spectrum from Fig. \ref{fig:nma-2DRamanOprobe} on an expanded scale,
together with several 1D horizontal and diagonal traces.  For
comparison, we also show the corresponding traces from the OON signal
using a blue dashed line.  These two signals differ only in the frequency
of the third pulse, the first and second pulses are the same. A trace
along the diagonal line $\Omega_1=\Omega_2$ is shown in panel
\textbf{i} and is very similar to the OO 1D-SXRS signal, the second
row, right panel from Fig. \ref{fig:nma-1DRaman}, with narrower
linewidths.  Although the $S_2$ (8.14 eV) is totally absent from the
OOO signal, which is consistent with the 1D results, this peak has
amplitude in the OON signal.  The matrix element of $\alpha_O$
connecting $S_0$ and $S_2,$ the square of which determines the
amplitude of this peak in the OO 1D spectrum, may be small but it must
be nonzero for this peak to show up in the OON signal.

A horizontal trace, with a constant $\Omega_2=8.95$ eV,  indicating that the system is in a $\ket{S_3}\bra{S_0}$ coherence during $\tau_2$,
is displayed in panel \textbf{ii}.  The diagonal peak is dominant here,  indicating a lack of off-diagonal peaks of
the first kind, those arising from diagrams $\S_{ii}$ and $\S_{iv}$.  The OOO and OON signals
are largely the same, with the latter being larger in magnitude. Panel \textbf{iii} shows a similar horizontal trace,
with $\Omega_2=8.14$ eV corresponding to $S_2$.  This trace is much larger in magnitude when the probe pulse is resonant with nitrogen core transitions, consistent with
the fact that the $S_2$ valence state is more strongly perturbed by a nitrogen than an oxygen core hole.
\par
The off-diagonal peaks seen in panels \textbf{ii} and \textbf{iii} result from the interference between diagrams
\begin{equation*}
\includegraphics[width = 4cm]{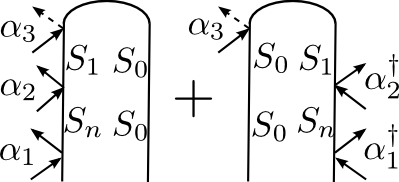}
\end{equation*}
where $S_n$ is the state giving rise to the $\Omega_1$ frequency.  The
$(\Omega_1,\Omega_2)=(8.95\, \textrm{eV},6.91\, \textrm{eV})$ peak in
panel \textbf{iii} depends nonlinearly on the matrix element
$\alpha_{O;S_3S_1}$.  Only off-diagonal peaks carry information about
the polarizability \emph{between} valence excited states.  The
diagonal peaks with contributions from all four diagrams in
Fig. \ref{fig:2Dramandiags} are insensitive to these quantities.  The
diagonal peak at $(\Omega_1,\Omega_2)=(8.95\, \textrm{eV},8.95\,
\textrm{eV})$ in panel \textbf{ii}, for example, depends on products
like $\alpha_{O;S_0S_3}\alpha_{O;S_3S_3}\alpha_{O;S_3S_0}$ where the
polarizability between the ground and valence excited states, which
determine the linear SXRS signal, are multiplied by a diagonal element
of the polarizability, seen to be much larger than off-diagonal
elements in Fig. \ref{fig:nma-2Dalphas}.

Panels \textbf{iv}, \textbf{v}, and \textbf{vi} show diagonal traces where $\Omega_2$ is equal to the difference between two valence excitation frequencies.
The peaks in panel \textbf{iv} result from interference of the following diagrams
\begin{equation*}
\includegraphics[width = 4.3cm]{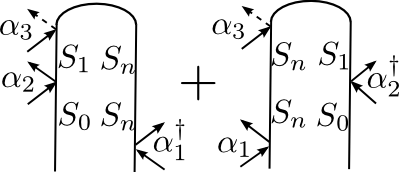}.
\end{equation*}
In panel \textbf{iv} the signal is much weaker with a nitrogen probe
than with an oxygen probe though they are similar in shape.  This is contrast with the two-color 1D results from Fig. \ref{fig:1DRaman}, where the $S_1$ peak is much stronger with a nitrogen probe than with an oxygen probe.
\par
Panels \textbf{ii} and \textbf{v} both show signals where the second pulse leaves the system in the $S_3$ state,
however they do so using the two different types of diagrams discussed above.  The
fact that these spectra are so different in appearance demonstrates the wealth of information available using 2D-SXRS.

\section{Concluding Remarks}
\label{sec:concluding}
We have calculated and analyzed multidimensional stimulated x-ray
Raman spectroscopy signals at the N and O K-edge.  The signals are
sensitive to the order in which the atom-specific x-ray excitations are
used to perturb and probe the valence electron configuration.  The
relative intensities of cross peaks could serve as a diagnostic tool
for the comparison of different levels of electronic structure theory.
By comparing frequency-domain RIXS and time-domain 1D-SXRS signals we
show that the ability to probe electronic dynamics at different atomic
centers is an advantage of the time-domain technique.  2D-SXRS offers
a greater number of pulse combinations, and greater control over the
preparation and measurement of electronic wavepackets.

The 1D-SXRS signal can be decomposed into the overlap of a doorway
with a window wavepacket in Hilbert space.  As discussed in
Sec. \ref{sec:2dsxrs}, the 2D-SXRS experiment can be written as the
sum of two valence wavepacket overlaps.  More detailed analysis of
this signal is needed to relate these terms to valence electronic
dynamics.  Stimulated Raman techniques in the x-ray frequency range
will allow many new and interesting measurements to be
performed. 2D-SXRS provides a new window into the correlated valence
electronic dynamics made possible by new x-ray sources.

\section*{Acknowledgments}
The support of the Chemical Sciences, Geosciences and Biosciences
Division, Office of Basic Energy Sciences, Office of Science,
U.S. Department of Energy is gratefully acknowledged. We also gratefully 
acknowledge the support of the National Science Foundation (Grant CHE-1058791), 
and the National Institutes of Health (Grant GM-59230).  We also wish to thank 
the anonymous reviewers whose suggestions
greatly improved the presentation of this paper.

\appendix
\section{The Effective Transition Polarizability for Gaussian Pulses}
\label{sec:derivation}
The effective polarizability has a simple form when the pulses have
Gaussian envelopes.  Assume the pulse envelopes are
\begin{equation}
 \mathcal{E}_j(t) = e^{-t^2/2\sigma_j^2-\phi_j},
\end{equation}
where $\sigma_j$ is the temporal pulse width (equal to intenstity FWHM
divided by $2 \sqrt{\ln 2}$), which determines the bandwidth of the
$j^\textrm{th}$ pulse.  Inserting this definition into the time-domain
expression for the polarizability (Eq. \ref{eq:13}), we get
\begin{equation}\begin{split} \alpha_{j;g'g''} =& -i\sum_e V_{g' e}V_{eg''} \int _{-\infty }^{\infty }d\tau _2\int _{-\infty }^{\tau _2}d\tau _1 \\ & \times
 \exp \left(-\frac{\tau_2^2}{2\sigma_j^2}-\frac{\tau_1^2}{2\sigma_j^2}+i \Delta^j_{eg'}\tau _2-i\Delta^j_{eg''}\tau _1\right).
 \end{split}\end{equation}
%First we switch the order of integration
%\begin{equation} \alpha_{x y}^{j k} = \sum_b V_{x b}V_{b y} e^{-\phi_k+\phi_j}\int _{-\infty }^{\infty }d\tau _1\int _{\tau_1 }^{\infty}d\tau _2
% \exp \left(-\frac{\tau_2^2}{2\sigma_j^2}-\frac{\tau_1^2}{2\sigma_k^2}+i \Delta^j_{eg'}\tau _2-i\Delta^j_{eg''}\tau _1\right)
%\end{equation}
Defining $\tau = \tau_2-\tau_1$ allows us to write
%\begin{equation}\begin{split} \alpha_{x y}^{j k} &= \sum_b V_{x b}V_{b y} e^{-\phi_k+\phi_j} \\ & \times
% \int _{-\infty }^{\infty }d\tau _1\exp \left(-\frac{\tau _1^2}{2\sigma _k^2}-i \Delta^j_{eg''}\tau _1\right)\int _0^{\infty }d\tau  \exp \left(-\frac{\left(\tau +\tau _1\right){}^2}{2\sigma _j^2}+i \Delta^j_{eg'}\left(\tau +\tau _1\right)\right)
%\end{split}\end{equation}
%Switching the order of integration once again gives
\begin{equation}\begin{split} \alpha_{j;g'g''} &= -i\sum_e V_{g'e}V_{eg''}
 \times\int _0^{\infty }d\tau  \exp \left(-\tau ^2/2\sigma _j^2+i \Delta^j_{eg'}\tau \right) \\
 & \times \int _{-\infty }^{\infty }d\tau _1\exp \left(-(\tau _1^2+\tau \tau_1)/\sigma _j^2-i \left(\Delta^j_{eg''}-\Delta^j_{eg'}\right)\tau_1\right)
\end{split}\end{equation}
The $\tau_1$ integral can be easily done by switching to polar
coordinates, and invoking the identity
\begin{equation}
 \int _{-\infty }^{\infty }dt e^{-t^2} = \sqrt{\pi}
\end{equation}
to get
\begin{equation}\label{eq:alphagausswithintdef}
\begin{split} \alpha_{j;g'g''} &= -i\sum_e V_{g'e}V_{eg''}  \sqrt{\pi}\sigma_j \exp \left(-(\Delta^j_{eg''}-\Delta^j_{eg'})^2\sigma _j^2/4\right) \\
& \times \int _0^{\infty }d\tau \exp\left(-\tau ^2/4\sigma _j^2+i \left(\Delta^j_{eg'}+\Delta^j_{eg''}\right)\tau/2  \right).
\end{split}\end{equation}
This integral can be written in terms of the complementary error
function,
\begin{equation}\label{eq:erfcdef}
 \int _{z }^{\infty }dt e^{-t^2} = \frac{\sqrt{\pi}}{2} \textrm{erfc}(z)
\end{equation}
Substituting Eq. \ref{eq:erfcdef} into Eq. \ref{eq:alphagausswithintdef} gives
\begin{equation} \begin{split} \alpha_{j;g'g''} =&  -i\sum_b V_{g'e}V_{eg''} \pi\sigma_j^2 \\ & \times
\exp \left(-\sigma_j^2\frac{(\Delta^j_{eg''}-\Delta^j_{eg'})^2+(\Delta^j_{eg''}+\Delta^j_{eg'})^2}{4} \right) \\ & \times
\text{erfc}\left(-i\sigma_j\frac{\Delta^j_{eg''} +\Delta^j_{eg'}}{2}\right), \end{split} \end{equation}
which can be simplified further to give Eq. \ref{eq:alpha}.

\section{Rotationally averaged all-parallel 2D-SXRS signal}
\label{sec:rotationalavg}

Here we give the all-parallel 2D-SXRS signal as a tensor contraction
over the polarization dependent expression
\begin{equation}\label{eq:2dramanstottens}
\S^{2D}(t_1,t_2) = \Im [\S_i(t_1,t_2) + \S_{ii}(t_1,t_2)
+ \S_{iii}(t_1,t_2) + \S_{iv}(t_1,t_2)],
\end{equation}
where $\S^{2D}$ is the sum of four terms (see Fig. \ref{fig:2Dramandiags})
\begin{eqnarray}\label{eq:sidef}
\S_{i}(t_1,t_2) &=& -
\langle
\al{2}{\dagger}{}(t_2)
\al{3}{}{}(t_2+t_1)
\al{1}{}{}(0)
\rangle \nonumber \\
&=& - \sum_{g' g''}
\al{2}{\nu_3 \nu_4 *}{g g''}
\al{3}{\nu_6 \nu_5}{g'' g'}
\al{1}{\nu_2 \nu_1}{g' g} \nonumber \\ & & \times
e^{-i (\epsilon_{g'}- i\gamma_{g'}) (t_1 + t_2) }
e^{+i (\epsilon_{g''} + i\gamma_{g''}) t_2 }
\end{eqnarray}
\begin{eqnarray}\label{eq:siidef}
\S_{ii}(t_1,t_2) &=&
\langle
\al{1}{\dagger}{}(0)
\al{2}{\dagger}{}(t_1)
\al{3}{}{}(t_1+t_2)
\rangle
\nonumber \\
&=&  \sum_{g' g''}
\al{1}{\nu_1 \nu_2 *}{g g'}
\al{2}{\nu_3 \nu_4 *}{g' g''}
\al{3}{\nu_6 \nu_5}{g'' g} \nonumber \\ & & \times
e^{+i (\epsilon_{g'} + i\gamma_{g'}) t_1 }
e^{+i (\epsilon_{g''} + i\gamma_{g''}) t_2 }
\end{eqnarray}
\begin{eqnarray}\label{eq:siiidef}
\S_{iii}(t_1,t_2) &=& -
\langle
\al{1}{\dagger}{}(0)
\al{3}{}{}(t_1+t_2)
\al{2}{}{}(t_1)
\rangle
 \nonumber \\
&=& - \sum_{g' g''}
\al{1}{\nu_1 \nu_2 *}{g g'}
\al{3}{\nu_6 \nu_5}{g' g''}
\al{2}{\nu_4 \nu_3}{g'' g} \nonumber \\ & & \times
e^{+i (\epsilon_{g'} + i\gamma_{g'}) (t_1+t_2) }
e^{-i (\epsilon_{g''}- i\gamma_{g''}) t_2 }
\end{eqnarray}
\begin{eqnarray}\label{eq:sivdef}
\S_{iv}(t_1,t_2) &=&
\langle
\al{3}{}{}(t_2+t_1)
\al{2}{}{}(t_1)
\al{1}{}{}(0)
\rangle
\nonumber \\
&=& \sum_{g' g''}
\al{3}{\nu_6 \nu_5}{g g''}
\al{2}{\nu_4 \nu_3}{g'' g'}
\al{1}{\nu_2 \nu_1}{g' g} \nonumber \\ & & \times
e^{-i (\epsilon_{g'}- i\gamma_{g'}) t_1 }
e^{-i (\epsilon_{g''}- i\gamma_{g''}) t_2 }.
\end{eqnarray}
The parallel signal is found by contracting the tensor signal
$\S^{2D}_{\nu_1 \dots \nu_6}(t_1,t_2)$
\begin{equation}
  \S^{2D}_{\parallel}(t_1,t_2) = \sum_{\nu_1 \dots \nu_{6}}
  I^{\parallel}_{\nu_1 \dots \nu_6}
  \S^{2D}_{\nu_1 \dots \nu_6}(t_1,t_2)
\end{equation}
with the 12th-order isotropic
tensor\cite{andrews_three-dimensional_1977}
\begin{eqnarray}
  I^{\parallel}_{\nu_1 \nu_2 \nu_3 \nu_4 \nu_5 \nu_6} &=& \frac{1}{105} \left(
 \delta_{\nu_1 \nu_2} \delta_{\nu_3 \nu_4} \delta_{\nu_5 \nu_6}
 +
 \delta_{\nu_1 \nu_2} \delta_{\nu_3 \nu_5} \delta_{\nu_4 \nu_6}
 \right.
 \nonumber \\
 &+& \quad
 \delta_{\nu_1 \nu_2} \delta_{\nu_3 \nu_6} \delta_{\nu_4 \nu_5}
 +
 \delta_{\nu_1 \nu_3} \delta_{\nu_2 \nu_4} \delta_{\nu_5 \nu_6}
 \nonumber \\
 &+& \quad
 \delta_{\nu_1 \nu_3} \delta_{\nu_2 \nu_5} \delta_{\nu_4 \nu_6}
 +
 \delta_{\nu_1 \nu_3} \delta_{\nu_2 \nu_6} \delta_{\nu_4 \nu_5}
 \nonumber \\
 &+& \quad
 \delta_{\nu_1 \nu_4} \delta_{\nu_2 \nu_3} \delta_{\nu_5 \nu_6}
 +
 \delta_{\nu_1 \nu_4} \delta_{\nu_2 \nu_5} \delta_{\nu_3 \nu_6}
 \nonumber \\
 &+& \quad
 \delta_{\nu_1 \nu_4} \delta_{\nu_2 \nu_6} \delta_{\nu_3 \nu_5}
 +
 \delta_{\nu_1 \nu_5} \delta_{\nu_2 \nu_3} \delta_{\nu_4 \nu_6}
 \nonumber \\
 &+& \quad
 \delta_{\nu_1 \nu_5} \delta_{\nu_2 \nu_4} \delta_{\nu_3 \nu_6}
 +
 \delta_{\nu_1 \nu_5} \delta_{\nu_2 \nu_6} \delta_{\nu_3 \nu_4}
 \nonumber \\
 &+& \quad
 \delta_{\nu_1 \nu_6} \delta_{\nu_2 \nu_3} \delta_{\nu_4 \nu_5}
 +
 \delta_{\nu_1 \nu_6} \delta_{\nu_2 \nu_4} \delta_{\nu_3 \nu_5}
 \nonumber \\
 &+& \quad
 \left.
 \delta_{\nu_1 \nu_6} \delta_{\nu_2 \nu_5} \delta_{\nu_3 \nu_4}
 \right).
\end{eqnarray}

\end{document}